\documentclass[lettersize,journal]{IEEEtran}
\usepackage{amsmath,amsfonts}
\usepackage{amssymb}
\usepackage{algorithmic}
\usepackage{algorithm}
\usepackage{array}
\usepackage[version=4]{mhchem}
\usepackage[caption=false,font=normalsize,labelfont=sf,textfont=sf]{subfig}

\usepackage{tabularx}
\usepackage{array}
\usepackage{pdflscape}
\usepackage{hyperref}
\usepackage{booktabs}
\usepackage{multirow}
\usepackage{textcomp}
\usepackage{stfloats}
\usepackage{url}
\usepackage{verbatim}
\usepackage{siunitx}
\usepackage{graphicx}
\usepackage{cite}
\title{AeroMig: Multi-modal Inverse-Gamma modeling for aerosol particle number size distributions}

\author{Abdur Rahman, 
        Juha Kangasluoma, 
        Santtu Mikkonen, 
        Tareq Hussein, 
        Tuukka Pet\"{a}j\"{a}, \\
        Sasu Tarkoma,
        and~Martha Arbayani Zaidan%

\thanks{A. Rahman and M.A. Zaidan are with Department of Computer Science and Institute for Atmospheric and Earth System Research, University of Helsinki, 00014 Helsinki, Finland (e-mails: abdur.rahman@helsinki.fi; martha.zaidan@helsinki.fi).}%
\thanks{S. Mikkonen is with Department of Environmental and Biological Sciences, Faculty of Science, Forestry and Technology, University of Eastern Finland, 70211 Kuopio, Finland (e-mail: santtu.mikkonen@uef.fi)}%
\thanks{J. Kangasluoma, T. Hussein, and T. Pet\"{a}j\"{a} are with Institute for Atmospheric and Earth System Research, University of Helsinki, 00014 Helsinki, Finland (e-mails: juha.kangasluoma@helsinki.fi; tareq.hussein@helsinki.fi; tuukka.petaja@helsinki.fi)}%
\thanks{S. Tarkoma is Department of Computer Science, University of Helsinki, 00014 Helsinki, Finland (e-mail: sasu.tarkoma@helsinki.fi).}%
}
\date{January 2025}

\begin{document}

\maketitle

\begin{abstract}
Aerosol particles play a critical role in air quality, human health, and climate processes, making their accurate characterization essential. One common way to represent aerosol populations is through particle number size distributions (PNSDs), which describe the concentration of particles across different mobility diameters. However, these distributions are often highly skewed, heavy-tailed, and multi-modal, posing significant challenges for conventional fitting approaches.
In this study, we propose a robust parametric framework (AeroMiG) for modeling PNSDs using Inverse-Gamma mixture distributions. 
The method represents observed size distributions as a superposition of multiple Inverse-Gamma components, enabling flexible modeling of asymmetric and heavy-tailed structures that are not well captured by widely-used Log-Gaussian based approaches.
The performance of the proposed method is evaluated using real-world data from three measurement stations.
Results are assessed using multiple criteria, including predictive accuracies, 
computational efficiency, and a composite score summarizing overall model quality. 
Comparative analysis against a widely used Aerosol Multi-mode Log-Gaussian (AeroMG) demonstrates that the AeroMiG framework consistently achieves improved fitting accuracy, lower error, and faster computation.
The proposed approach provides a reliable and computationally efficient solution for modeling complex aerosol size distributions, with potential applications in real-time air quality monitoring and data-driven environmental analysis.

\end{abstract}

\section{Introduction}

Atmospheric science is a fundamental discipline for understanding the physical, chemical, and dynamical processes that govern the Earth’s atmosphere. It plays a central role in climate research, air quality assessment, and environmental monitoring by providing the scientific basis for interpreting interactions between natural systems and anthropogenic activities \cite{seinfeld_atmospheric_2016, intergovernmental_panel_on_climate_change_ipcc_climate_2023, masson2021climate}. 

One of the most critical challenges in atmospheric science is the study of climate change, where the atmosphere acts as a key medium for energy balance and radiative forcing \cite{hinds_aerosol_2022, li_scattering_2022}. Greenhouse gases and aerosol particles influence the Earth’s radiation budget through absorption, scattering, and cloud interactions, thereby affecting temperature, precipitation patterns, and extreme weather events \cite{akinyoola_review_2024}. In particular, aerosols remain one of the largest sources of uncertainty in climate projections due to their complex spatial variability and interactions with clouds and radiation \cite{intergovernmental_panel_on_climate_change_ipcc_climate_2023, stocker_climate_2014}. Advances in atmospheric measurement and data analysis are critical for reducing uncertainties in climate models, improving air quality forecasts, and enhancing our ability to respond to environmental challenges in a changing climate \cite{Karpatne2019}.

Fine particulate matter, particularly particles with diameters below 2.5~$\mu$m (PM$_{2.5}$), has been strongly associated with adverse health outcomes, including respiratory and cardiovascular diseases, and premature mortality \cite{Du2021}, while atmospheric aerosols also play a critical role in the climate system by directly scattering and absorbing solar radiation and indirectly modifying cloud formation, lifetime, and albedo \cite{seinfeld_atmospheric_2016, masson2021climate, pope_iii_lung_2002}. 

A key representation of aerosol populations is the PNSD, which describes the concentration of particles as a function of their mobility diameter. PNSDs provide detailed insight into aerosol sources, transformation processes such as nucleation, coagulation, and condensation, and the evolution of atmospheric particle populations \cite{hinds_aerosol_2022, whitby_physical_1978}.

Measured PNSDs are often noisy, highly skewed, and influenced by multiple overlapping physical processes such as nucleation, condensation, and coagulation, which make direct interpretation difficult and necessitate the use of parametric fitting approaches to extract meaningful information about underlying aerosol populations \cite{seinfeld_atmospheric_2016, hinds_aerosol_2022}.

Fitting algorithms provide a systematic way to approximate observed distributions using a combination of analytical functions, allowing the decomposition of complex PNSDs into interpretable modes associated with distinct sources or transformation mechanisms \cite{whitby_physical_1978}. However, the fitting process is inherently challenging due to the nonlinear and multi-parameter nature of the problem, which often leads to multiple local minima and parameter identifiability issues. Traditional optimization methods may struggle to provide stable and consistent solutions, particularly when the data exhibit strong asymmetry and heavy-tailed behavior.

To address these challenges, advanced optimization strategies are required to ensure reliable parameter estimation and global convergence. Evolutionary algorithms, such as Differential Evolution, offer a robust alternative by performing population-based global search, reducing sensitivity to initialization, and improving the likelihood of identifying optimal or near-optimal solutions in complex search spaces \cite{storn_differential_1997}. These properties make such algorithms particularly suitable for fitting aerosol size distributions, where both model flexibility and numerical stability are essential. 
Some different models have been used for modeling aerosol size particles. 
The prevailing approach relies on lognormal modal decomposition, where aerosol populations are represented as a superposition of modes associated with nucleation, Aitken, accumulation, and coarse particles \cite{whitby_modal_1997, hinds_aerosol_2022, Hussein2005}. This framework is widely used because of its physical interpretability and consistency with classical aerosol theory.

The AeroMG (Aerosol Multi-mode Log-Gaussian) 

algorithm has become a standard tool for estimating modal parameters through nonlinear least-squares optimization \cite{Hussein2005}. While effective in many applications, AeroMG and related approaches depend on assumptions of lognormality and often require careful initialization and predefined mode structures \cite{zhu_comprehensive_2024, Hussein2005}. As a result, their performance can degrade when applied to highly variable, asymmetric, or heavy-tailed distributions frequently observed in real atmospheric measurements. 
To improve flexibility, alternative statistical approaches such as Gaussian mixture models and nonparametric density estimation have been explored \cite{seinfeld_atmospheric_2016}. Although these methods can better capture multi-modal behavior, they often lack clear physical interpretability and may suffer from instability or overfitting in the presence of noisy data. More recently, global optimization techniques, including evolutionary algorithms, have been introduced to address the non-convex nature of the fitting problem and reduce sensitivity to initialization \cite{storn_differential_1997}. However, their integration with physically meaningful distribution models for aerosol data remains limited. Despite these developments, a key limitation persists such that existing approaches struggle to simultaneously capture heavy-tailed behavior, maintain physical interpretability, and ensure stable parameter estimation. This gap motivates the development of alternative modeling frameworks that combine flexible distributional forms with robust global optimization strategies.

To address these limitations, this paper introduces AeroMiG (Aerosol Multi-mode Inverse-Gamma), a framework explicitly engineered to capture the highly skewed, heavy-tailed, and asymmetric structures of ultrafine particles in aerosol PNSDs. This work delivers several key contributions. First, by incorporating a stability-driven objective function, the framework resolves common convergence issues while maintaining strict physical interpretability, enabling the reliable tracking of nucleation representation and multi-mode dynamics. Second, the model is validated using extensive real-world datasets across multiple stations, encompassing a 3-mode configuration spanning 15 years and a 5-mode configuration covering full seasonal cycles. Third, comprehensive evaluations demonstrate that AeroMiG consistently outperforms the state-of-the-art AeroMG framework in both fitting accuracy and computational efficiency. Finally, the model's performance is uniquely benchmarked across distinct NPF event groups to evaluate its robustness under varying atmospheric conditions.

\section{Measurement Sites and Data}

This section describes the primary datasets utilized in this study, specifically PNSD data, followed by an overview of the SMEAR (Station for Measuring Ecosystem-Atmosphere Relations) stations where these measurements and relevant variables were collected.

\subsection{Aerosol Measurement data}

PNSD were measured using a differential mobility particle sizer (DMPS), which determines size-resolved aerosol number concentrations based on electrical mobility diameter. The system combines bipolar charge neutralization, size classification using a differential mobility analyser, and detection via a condensation particle counter, enabling continuous measurements across the nucleation, Aitken, and accumulation modes.

\begin{figure}[ht]
    \centering \includegraphics[width=0.5\textwidth, height=0.4\textheight, keepaspectratio]{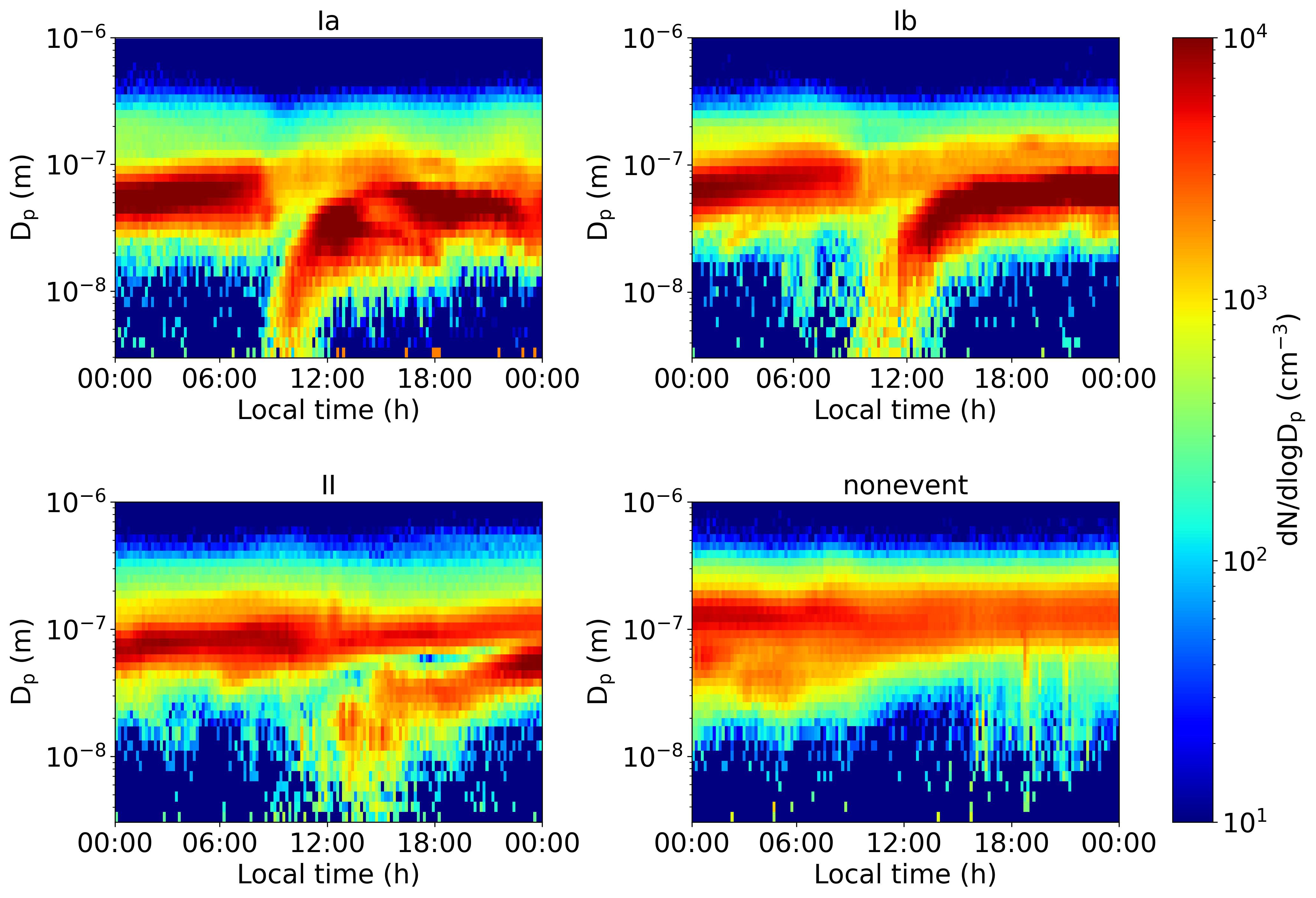}
    \caption{Example PNSD plots from DMPS measurements. Axes show time (x) and particle size in nm (y); color represents particle concentration in  cm$^{-3}$.}
    \label{NPF}
\end{figure}

A critical atmospheric phenomenon observed via PNSD data is New Particle Formation (NPF). In this study, NPF events are classified following the widely adopted methodology of Dal Maso et al. \cite{dal2005formation}, which categorizes days based on the clarity of the nucleation mode appearance and the persistence of subsequent particle growth. As illustrated in Fig. \ref{NPF}, these events are structured into four distinct classes \cite{Zaidan2018}: Class I events (subdivided into Ia and Ib) exhibit robust formation and distinct growth tracks of new particles; Class II events indicate weaker or less well-defined growth trajectories; and nonevent days correspond to the absolute absence of detectable nucleation. 
In this work, the performance of AeroMiG and AeroMG is evaluated and compared across these distinct NPF event groups to uncover category-specific behavioral differences.

\subsection{SMEAR Stations}
\begin{figure}[ht]
    \centering \includegraphics[width=0.49\textwidth, height=\textheight, keepaspectratio]{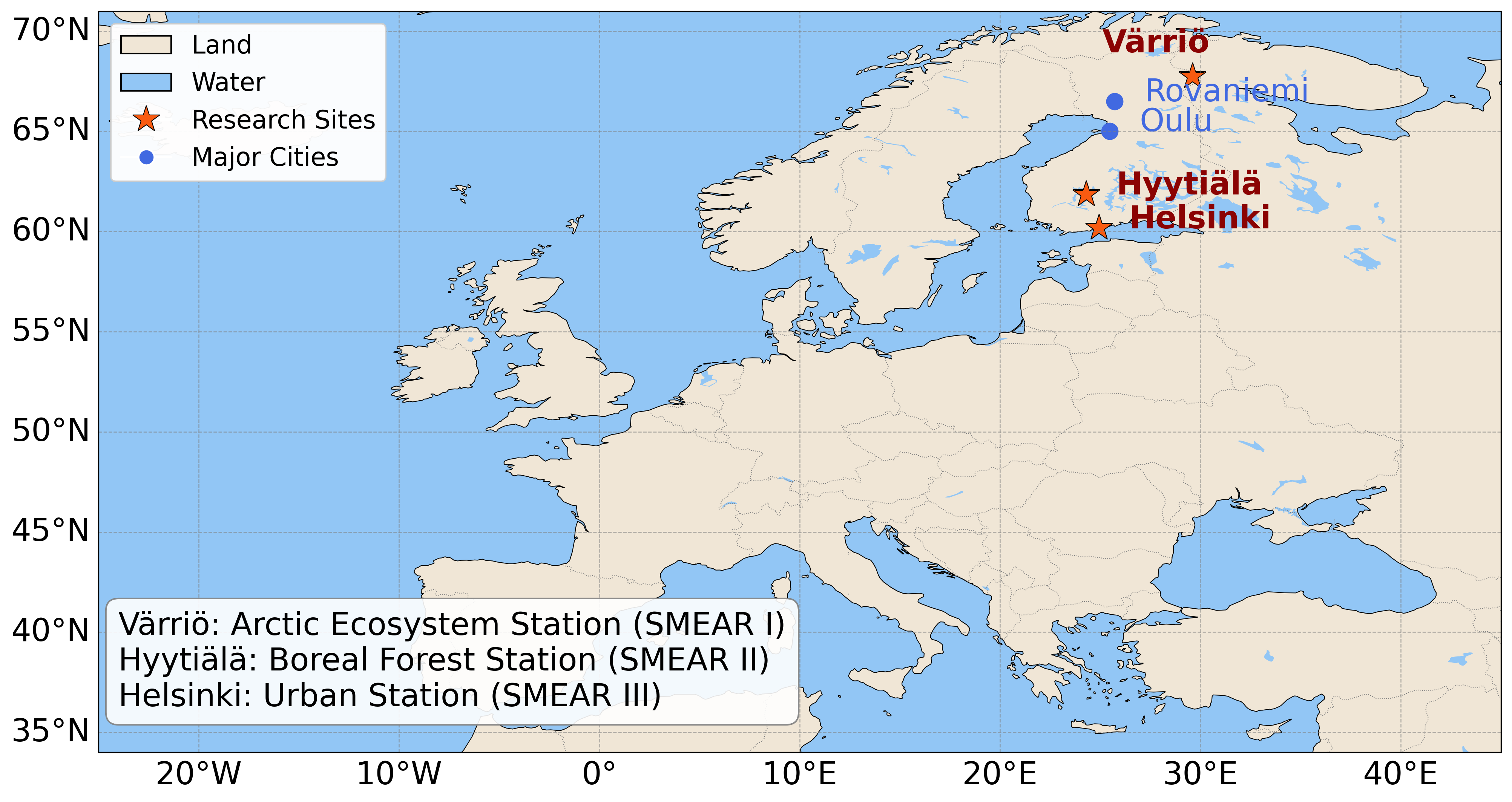}
    \caption{Geographic distribution of the SMEAR stations providing data for this study.}
    \label{domain}
\end{figure}

This subsection describes the three SMEAR stations where the PNSD data were collected as shown in Fig. \ref{domain}. 

\subsubsection{SMEAR I}

The Värriö Subarctic Research Station (SMEAR I; $67.75^{\circ}$N, $29.61^{\circ}$E) is located $\sim 120$ \text{km} north of the Arctic Circle in the northern boreal zone (Fig. \ref{domain}), near the timberline within a strictly protected nature reserve ($125$ \text{km}$^2$)\cite{hari_station_2013}. Due to its pristine conditions and minimal anthropogenic influence, the primary purpose of SMEAR I is the continuous monitoring and study of subarctic aerosol formation, ion dynamics, and biosphere–atmosphere exchange \cite{ahti_vegetation_1968, kyro_trends_2014}. General meteorological variables are collected via a 16 m meteorological mast, alongside outdoor tree chamber measurements for photosynthesis. Aerosol properties, including PNSDs, and trace gases are continuously sampled and measured using specialized analyzers housed inside a dedicated, climate-controlled measurement hut on site \cite{kyro_trends_2014}.

\subsubsection{SMEAR II}

The Hyytiälä measurement station (SMEAR II; $61.95^{\circ}$N, $24.28^{\circ}$E) is located $181$ \text{m} above sea level in a non-urbanized boreal forest environment (Fig. \ref{domain}), approximately $190$ \text{km} northwest of Helsinki \cite{hari2005}. As one of the most comprehensive in-situ observation sites globally, its primary purpose is the long-term, continuous monitoring of ecosystem–atmosphere interactions, atmospheric chemistry, and NPF studies \cite{hari2005, kulmala_formation_2004}. Since $1996$, the station's extensive instrumentation has continuously measured general variables including micrometeorology, weather conditions, and cloud remote sensing. PNSDs and trace gas concentrations are continuously sampled and tracked using specialized online analyzers deployed across the station's ground-based facilities and dedicated measurement towers \cite{hari2005, Uusitalo2021}.

\subsubsection{SMEAR III}

The SMEAR III station ($60.20^{\circ}$N, $24.96^{\circ}$E; $26$ \text{m} above sea level) is located in Kumpula (Fig. \ref{domain}), approximately $3$ \text{km} northeast of Helsinki's center, situated within a heterogeneous urban background environment characterized by a mixture of infrastructure, traffic corridors, and vegetated surfaces \cite{hussein_urban_2004, hussein_meteorological_2006}. The primary purpose of SMEAR III is to characterize urban aerosol processes, meteorological conditions, and biosphere–atmosphere interactions under diverse anthropogenic and urban source influences \cite{hussein_meteorological_2006}. General environmental variables are measured from a $31$ \text{m} tower situated on a rocky hill. PNSDs are continuously sampled using a DMPS housed in a nearby measurement container \cite{hussein_meteorological_2006}. While measurements originally began at a nearby site in Siltavuori ($1997\text{–}2001$) before relocating to Kumpula, an instrumental upgrade on $9$ March $2003$ successfully extended the measurable PNSD range from $7\text{–}500$ \text{nm} to $3\text{–}1000$ \text{nm} without introducing discontinuities in the long-term data characterization \cite{aalto_physical_2001, hussein_urban_2004}.

\section{Methods}

This section provides detailed descriptions of the AeroMG and AeroMig models, alongside the performance metrics employed to compare them.

\subsection{AeroMG - Aerosol Multi-mode Log-Gaussian}

Atmospheric PNSDs are commonly represented as a superposition of multiple Log-Gaussian (lognormal) modes, reflecting contributions from distinct aerosol processes. For a given particle diameter $D_p$, the modeled PNSD is

\begin{equation}
\hat{n}(D_p)
=
\sum_{k=1}^{K}
\frac{N_k}{\sqrt{2\pi}\, D_p\ln \sigma_k}
\exp\left[
-\frac{(\ln D_p - \ln D_{pg,k})^2}{2(\ln \sigma_k)^2}
\right],
\end{equation}
where $K$ is the number of modes, $D_{pg,k}$ is the geometric mean diameter of mode $k$, $\sigma_k$ is the geometric standard deviation of mode $k$, $N_k$ is the total number concentration of mode $k$. The full parameter vector is therefore
\[
\boldsymbol{\theta}
=
\{ D_{pg,1}, \ldots, D_{pg,K},\;
\sigma_1, \ldots, \sigma_K,\;
N_1, \ldots, N_K \}.
\]
Given a measured PNSDs $\{(D_{p,i}, n_i) \}_{i=1}^{M}$ consisting of $M$ measurement bins, where $n_i$ represents the measured concentration value at diameter $D_{p,i}$, the fitting procedure minimizes the mean squared error (MSE) between the measured and modeled distributions:

\begin{equation}
\text{MSE}
=
\frac{1}{M}
\sum_{i=1}^{M}
\left[
n_i - \hat{n}(D_{p,i}; \boldsymbol{\theta})
\right]^2.
\end{equation}
The fitting parameters $\boldsymbol{\theta}$ are obtained by nonlinear optimization subject to physical constraints:
\[
D_{pg,k} > 0, \quad \sigma_k > 1, \quad N_k > 0.
\]
After the initial fit with a maximum number of modes $K_{\max}$, the algorithm automatically reduces the number of modes through two steps:

\subsubsection{Elimination of Invalid Modes}

Modes with negligible number concentration or physically invalid parameters are removed:
\begin{equation}
N_k \leq \varepsilon \quad \Rightarrow \quad \text{mode } k \text{ removed},
\end{equation}
where $\varepsilon$ is a small numerical threshold.

\subsubsection{Mode Overlap Reduction}

Adjacent modes are tested for overlap using a tolerance-based criterion. If the reduction in fitting error after removing a mode is below a predefined tolerance ($TOL_K$), the simpler model is preferred:
\begin{equation}
\Delta \text{MSE} = \text{MSE}_{K-1} - \text{MSE}_K < TOL_K
\quad \Rightarrow \quad
K \leftarrow K - 1.
\end{equation}
This reduction is performed iteratively until removing further modes exceeds the error tolerance.

\subsection{AeroMiG- Aerosol Multi-mode Inverse-Gamma}

DMPS instruments quantify aerosol particle mobility diameters $D_p > 0$, producing right-skewed, heavy-tailed size distributions \cite{colbeck_aerosol_2014, hinds_aerosol_2022, seinfeld_atmospheric_2016}. In many atmospheric and combustion environments, the probability of observing extremely small particles declines sharply, whereas large particles occur with low but non-negligible probability \cite{whitby_modal_1997, curtius_nucleation_2006}. Such behavior is characteristic of inverse-scale processes, motivating the use of the Inverse-Gamma distribution as a parametric model for particle diameters.\\
Let $Y \sim Gamma(\alpha,\beta)$ with shape $\alpha>0$ and scale $\beta>0$. Now consider the random variable $Y = \frac{1}{D_p}$, then the distribution of $D_p$ is exactly the inverse of the gamma variable, which is Inverse-Gamma distribution. If $D_p = g(Y)$ where $g(y) = 1/y$, then $y = 1/D_p$ and the Jacobian $\left| \frac{dy}{dD_p} \right| = \frac{1}{D_p^2}$. Applying this transformation, we get
\begin{align}
f_{D_p}(D_p \mid \alpha, \beta) &= f_Y(1/D_p) \left| \frac{d}{dD_p} (1/D_p) \right| \\
&= \frac{\beta^\alpha}{\Gamma(\alpha)} \left(\frac{1}{D_p}\right)^{\alpha-1} \exp\Big(-\frac{\beta}{D_p}\Big) \cdot \frac{1}{D_p^2} \\
&= \frac{\beta^\alpha}{\Gamma(\alpha)} D_p^{-(\alpha+1)} \exp\Big(-\frac{\beta}{D_p}\Big), \quad D_p > 0.
\label{pdf}
\end{align}
This distribution is strictly supported on $(0, \infty)$ and exhibits a heavy right tail. Given a measured PNSDs $\{(D_{p,i}, n_i) \}_{i=1}^{M}$, the log-likelihood function is
\begin{align*}
\ln L(\alpha, \beta) &= \sum_{i=1}^M n_i \ln f(D_{p,i} \mid \alpha, \beta) \\
&= \left(\sum_{i=1}^M n_i\right) \big(\alpha \ln \beta - \ln \Gamma(\alpha)\big) \\
&- (\alpha + 1) \sum_{i=1}^M n_i \ln D_{p,i} - \beta \sum_{i=1}^M \frac{n_i}{D_{p,i}}
\end{align*}
Let $N' = \sum_{i=1}^M n_i$, then maximizing $\ln L$ with respect to $\beta$ yields a closed-form estimator conditional on 
$\alpha$ such that
\begin{equation}
\hat{\beta} = \frac{N' \alpha}{\sum_{i=1}^M n_i D_{p,i}^{-1}}
\end{equation}
The nonlinear equation for $\alpha$ becomes:
\begin{equation}
\ln \hat{\beta} - \psi(\alpha) - \frac{1}{N'} \sum_{i=1}^M n_i \ln D_{p,i} = 0
\end{equation}
where $\psi(\cdot)$ is the digamma function. This equation requires numerical solution. The mean and variance are:
\begin{align}
\mathbb{E}[D_p] &= \frac{\beta}{\alpha - 1}, \quad \alpha > 1,\\
\mathrm{Var}(D_p) &= \frac{\beta^2}{(\alpha - 1)^2 (\alpha - 2)}, \quad \alpha > 2.
\end{align}
The Probability Density Function (PDF) (Eq. \ref{pdf}) is commonly computed in $\log$-space such that

\begin{equation}
\ln f(D_p \mid \alpha, \beta) = \alpha \ln \beta - \ln \Gamma(\alpha) - (\alpha + 1) \ln D_p - \frac{\beta}{D_p},
\end{equation}
which preserves floating-point accuracy in high-performance numerical kernels.

Direct evaluation of $\beta^\alpha D_p^{-\alpha - 1}e^{-\beta / D_p}$ (Eq. \ref{pdf}) can be unstable when $\alpha$ is large, when $D_p$ is very small (leading to overflow in $D_p^{-\alpha - 1}$), or when $\beta/D_p$ is large (leading to underflow). Therefore, the density is commonly computed using log-space decompositions, especially when high-performance numerical kernels are required. The normalization constant can be written as
\begin{equation}
\frac{\beta^\alpha}{\Gamma(\alpha)} = \exp(\alpha \ln \beta - \ln \Gamma(\alpha))
\end{equation}
which avoids overflow in $\beta^\alpha$ and leverages the numerically stable computation of the logarithmic gamma function. Likewise, the power term is evaluated via
\begin{equation}
D_p^{-\alpha - 1} = \exp((-\alpha - 1) \ln D_p)
\end{equation}
to avoid under/overflow. Combining these terms, the entire probability density is evaluated through exponentiating the log-space formulation to preserve floating-point accuracy:
\begin{equation}
f(D_p \mid \alpha, \beta) = \exp(\alpha \ln \beta - \ln \Gamma(\alpha) - (\alpha+1) \ln D_p - \frac{\beta}{D_p})
\end{equation}
Because $\alpha$ controls the decay rate of large-diameter occurrence and $\beta$ sets a characteristic inverse scale, the distribution provides interpretable aerosol microphysical parameters.

\subsubsection*{Interpretability of AeroMiG} 
For the Inverse-Gamma distribution applied to DMPS mobility diameters 
$D_p$, the parameters $\alpha$ and $\beta$ have physically interpretable aerosol micro-physical roles. The shape parameter $\alpha$ primarily controls the heaviness of the right (upper) tail of the distribution. Small values of $\alpha (\alpha \lesssim 2)$ correspond to broad size variability and an increased probability of large, aged, or coagulated particles, whereas larger 
$\alpha$ values indicate a more size-homogeneous aerosol population dominated by ultra-fine particles with limited coarse-mode contribution. Thus, 
$\alpha$ provides a statistical indicator of aerosol aging, coagulation intensity, and source mixing.
The scale parameter $\beta$ determines the characteristic diameter of the distribution. Because the mode of the Inverse-Gamma distribution is

\begin{equation}
D_{p, mode} = \frac{\beta}{\alpha + 1}, \quad \alpha > 1,
\end{equation} 
larger $\beta$ values shift the particle-size distribution toward larger modal diameters, indicating an overall redistribution of particle mass toward larger sizes. While condensational growth affects particles across the entire size spectrum, such a shift in the mode may reflect the combined influence of growth processes, secondary formation, and source characteristics. Conversely, smaller $\beta$ values are associated with freshly nucleated or combustion-derived ultra-fine particles.
When $\alpha > 1$, the mean particle diameter is finite and equals \begin{equation}
\mathbb{E}[D_p] = \frac{\beta}{\alpha - 1},
\end{equation}
indicating that the ratio $\frac{\beta}{\alpha - 1}$ can be interpreted as the expected mobility diameter. For $\alpha \le 1$, the mean diverges, reflecting environments with highly intermittent coarse particles.
Taken together, $\alpha$ describes distribution sharpness and tail behavior, while $\beta$ establishes the dominant particle-size scale. Their combination provides a compact, interpretable parameterization of DMPS aerosol size structure, enabling comparison across sources, air masses, and atmospheric processes.

\subsection{Performance Metrics}

The fitting problem is inherently ill-posed and multi-modal, where multiple parameter combinations may yield similar residual errors. Relying solely on the Mean Squared Error (MSE) to assess model performance can therefore favor over-parameterized solutions that reproduce the data numerically but lack physical interpretability \cite{bishop_pattern_2016, huber_robust_2009, murphy_machine_2012, zhao_infovae_2019}. To achieve a balanced evaluation that accounts for both accuracy and model parsimony, a composite performance metric integrating five complementary statistical measures was adopted: MSE, Akaike Information Criterion (AIC), Bayesian Information Criterion (BIC), the coefficient of determination ${R}^2$, and adjusted ${R}^2$.

\paragraph{Mean Squared Error (MSE)} 
The MSE quantifies the average squared deviation between the measured and modeled PNSD across all particle diameter bins:

\begin{equation}
\text{MSE} = \frac{1}{N}\sum_{k=1}^{N}(y_k - \hat{y}_k)^2,
\end{equation}
where $y_k$ and $\hat{y}_k$ denote the measured and modeled particle concentrations at the $k$-th diameter bin, respectively. Thus, MSE provides a single scalar measure summarizing the overall discrepancy between the modeled and observed size distributions.\\
Lower MSE values indicate closer agreement between modeled and observed particle number concentrations. However, because MSE typically decreases with increasing model complexity, it may not reliably indicate the optimal number of modes without additional model selection criteria \cite{nyaku_comparison_2020, ren_prediction_2020, hastie_elements_2009, burnham_model_2010}.

\paragraph{Akaike Information Criterion (AIC)} 
The AIC introduces an information-theoretic penalty for additional model parameters:
\begin{equation}
\text{AIC} = 2v - 2\ln(L)
\end{equation}
where $v$ is the number of fitted parameters and $L$ is the maximum likelihood. When applying MSE as a fitting proxy, the negative log-likelihood is approximated as proportional to MSE under the assumption of approximately Gaussian residuals. This ensures that improvements in fit quality are justified by corresponding increases in model evidence rather than overfitting \cite{sugiura_further_1978, akaike_new_1974, burnham_model_2010}.

\paragraph{Bayesian Information Criterion (BIC)} 
Similar to AIC but with a stronger penalty for model complexity, the BIC is expressed as:
\begin{equation}
\text{BIC} = v\ln(n) - 2\ln(L)
\end{equation}
BIC tends to favor simpler models when data points ($n$) are limited, providing a safeguard against overfitting in the presence of noise or measurement uncertainty common in atmospheric datasets \cite{schwarz_estimating_1978, kass_bayes_1995, burnham_model_2010}.

\paragraph{Coefficient of Determination ($R^2$)} 
The coefficient of determination ($R^2$) quantifies how well the fitted model reproduces the observed PNSD across all diameter bins. Specifically, $R^2$ is computed by comparing the measured and modeled concentration values at each particle diameter $D_{p,k}$:

\begin{equation}
R^2 = 1 - \frac{\sum_{k=1}^{N} (y_k - \hat{y}_k)^2}{\sum_{k=1}^{N} (y_k - \bar{y})^2},
\end{equation}
where $y_k$ and $\hat{y}_k$ denote the measured and modeled particle concentrations at the $k$-th diameter bin, respectively, and $\bar{y}$ is the mean of the measured distribution. This formulation treats the PNSD as a discrete function sampled over 
$N$ diameter bins.\\
Thus, $R^2$ provides a single scalar measure summarizing the goodness-of-fit over the entire size distribution, with higher values indicating that the model effectively captures the structural variability of the PNSD \cite{gao_particle_2023, yang_comparison_2012}.

\paragraph{Adjusted $R^2$} 
Adjusted $R^2$ accounts for model complexity by correcting the coefficient of determination for the number of fitted parameters:

\begin{equation}
R^2_{\text{adj}} = 1 - \left(1 - R^2\right)\frac{N-1}{N-v-1},
\end{equation}
where $N$ is the number of diameter bins and $v$ is the number of fitted parameters. This metric prevents artificial inflation of $R^2$ as additional modes are introduced and favors parsimonious models that achieve high explanatory power with fewer components \cite{wherry_new_1931, kutner_applied_2005, yang_comparison_2012}.

\paragraph{Composite Score}
Due to the different scaling and boundedness of the performance metrics, direct comparison or aggregation is not meaningful. Therefore, each metric $f_m$ is transformed into a bounded score using an independent monotonic transformation $$S_m(f_m) \in (0, 1),$$ such that the resulting score preserves the ordering of model performance while mapping all metrics onto a common scale.
For metrics where smaller values indicate better performance (MSE, AIC, BIC), a decreasing logistic transformation is applied:

\begin{equation}
S_m(f_m) = \frac{1}{1 + \exp{({\delta} g(f_m))}}
\end{equation}

where ${\delta} > 0$ controls the steepness of the transformation and $g(f_m)$ is a compression function defined as $$g(f_m) = \log(1 + |f_m|).$$
This logarithmic compression reduces the dynamic range of large values while preserving the ordering of the metric.
Thus, for the error-based metrics,

\begin{equation}
S_v = \frac{1}{1 + \exp{({\delta}\log(1 + \text{MSE}))}}
\end{equation}

\begin{equation}
S_a = \frac{1}{1 + \exp{({\delta} \,\text{sign(AIC)}\log(1 + |\text{AIC}|))}}
\end{equation}

\begin{equation}
S_b = \frac{1}{1 + \exp{({\delta} \,\text{sign(BIC)}\log(1 + |\text{BIC}|))}}
\end{equation}
The coefficients of determination ${R}^2$ and $\text{adjusted} \ {R}^2$ are naturally bounded above by unity and therefore require no additional transformation. Their normalized scores are simply defined as
$$S_r = {R}^2$$
$$S_{ar} = {R}_{adj}^2.$$ The final composite score integrates these five normalized metrics using empirically determined weights:
\begin{equation}
S_c = \frac{
w_v S_v + w_a S_a + w_b S_b + w_{r} S_{r} + w_{ar} S_{ar}
}{
w_v + w_a + w_b + w_{r} + w_{ar}
}
\end{equation}
or equivalently
\begin{equation}
S_c = \frac{1}{\sum_j w_j} \sum_j w_j \, S_j,
\label{eq:composite}
\end{equation}
where $w_j$ denotes the weight assigned to the $j-\text{th}$ metric and $S_j$ its normalized score. The weights 
$$(w_v, w_a, w_b, w_r, w_{ar}) = (0.2, 0.15, 0.15, 0.25, 0.25)$$ 
assign equal total importance to error minimization (\text{MSE}, \text{AIC}, \text{BIC}) ($50\%$) and statistical robustness and explanatory power ($\mathrm{R}^2$, $\text{adjusted} \ \mathrm{R}^2$) ($50\%$). This balanced weighting strategy prevents over-parameterization while ensuring high fidelity to the underlying physical data structure \cite{burnham_model_2004}. The composite score provides a bounded scalar measure of overall model performance:
$$0 < S_c < 1.$$ 
The model with the highest composite score is selected as the best-fit representation of the measured PNSD.
This multi-criteria approach balances goodness-of-fit, model complexity, and explanatory power, ensuring that the selected solution achieves both high reconstruction fidelity and statistical parsimony. Such balanced evaluation is particularly important for atmospheric nucleation studies, where small numerical differences in fit quality can correspond to substantial differences in inferred particle formation and growth mechanisms.

\section{Results and Discussion}

This section presents a comparative analysis between the proposed AeroMig model and AeroMG \cite{Hussein2005, zhu_comprehensive_2024}. Furthermore, it discusses the practical advantages of AeroMig within various real-world contexts.

\subsection{Results}

This subsection evaluates the performance of the models using multiple metrics, including predictive accuracies which are quantified by Mean Square Error (MSE), $R^2$, and the composite score metric as well as computational efficiency. These evaluations were conducted using PNSD data collected across all SMEAR stations.

\subsubsection{Predictive Accuracies} 
\begin{figure}[htbp]
    \centering    \includegraphics[width=0.5\textwidth]{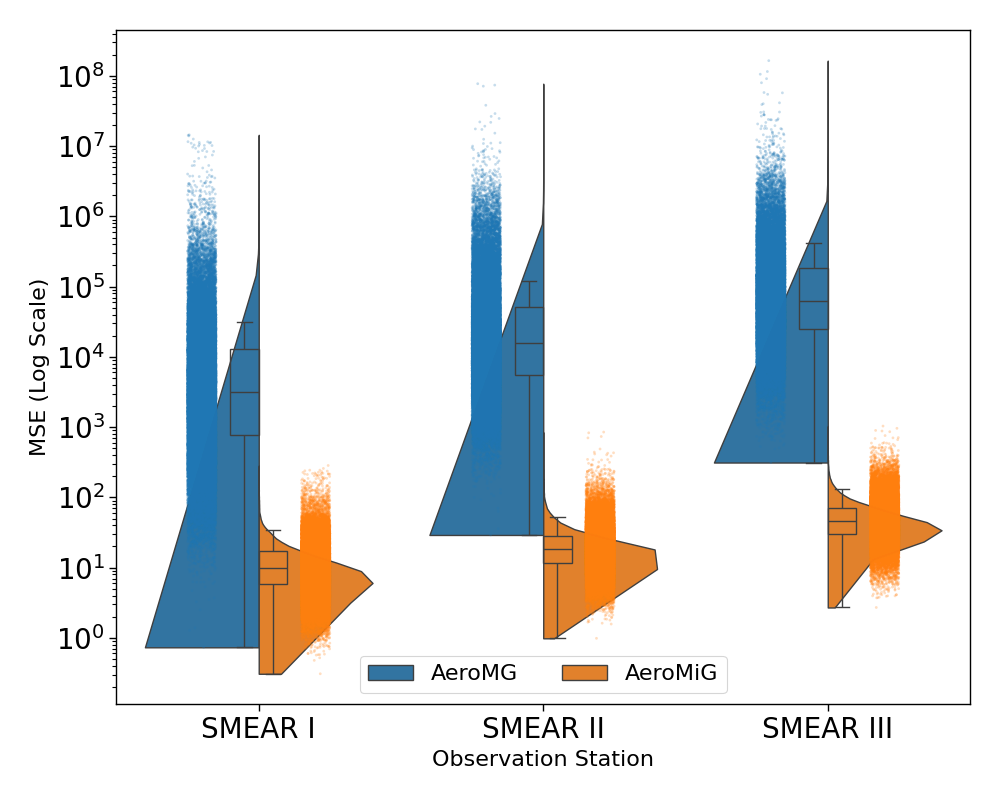}
    \caption{Raincloud plot of Mean Square Error (MSE) in different stations}
    \label{fig:MSE_all}
\end{figure}

\begin{figure}[htbp]
    \centering \includegraphics[width=0.5\textwidth]{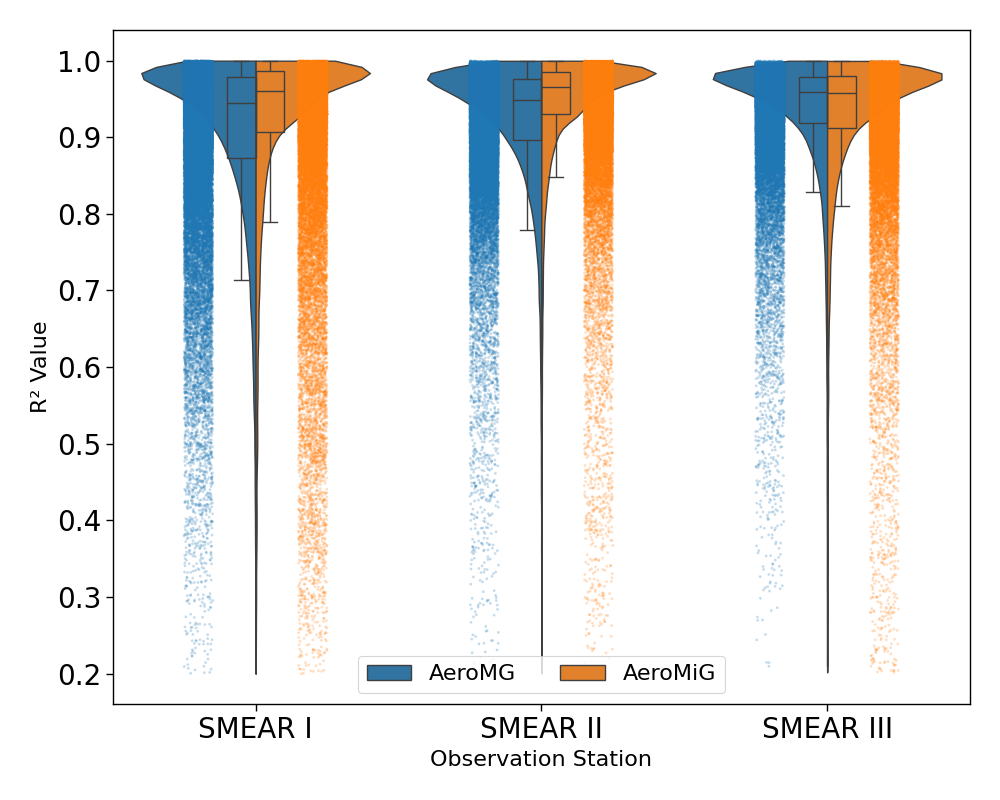}
    \caption{Raincloud plot of R-squared value in different stations}
    \label{fig:R2_all}
\end{figure}

\begin{figure}[htbp]
    \centering \includegraphics[width=0.5\textwidth]{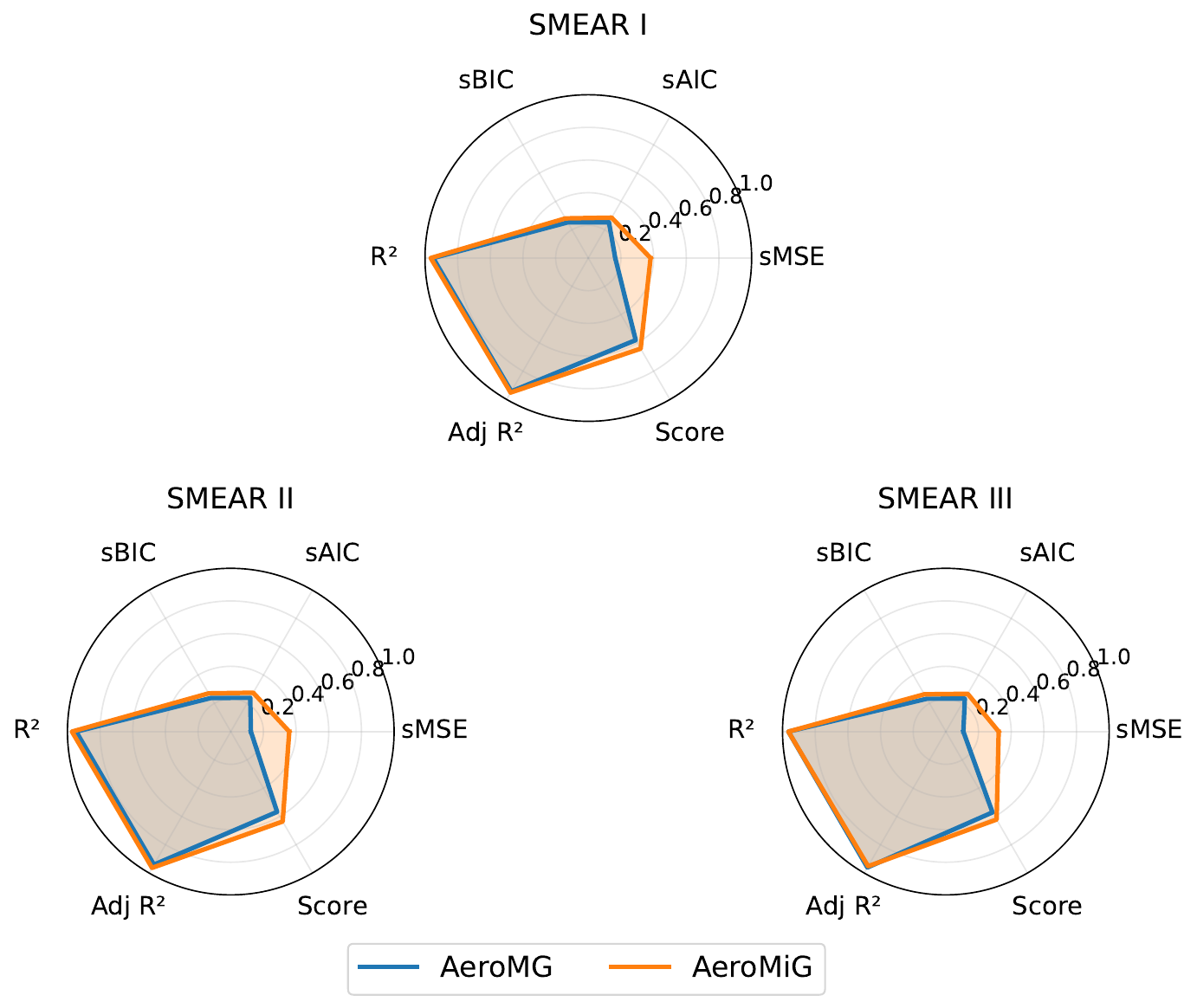}
    \caption{Polar-plot of performance matrix and their composite score in different stations}
    \label{fig:score}
\end{figure}

serve as performance metrics to quantify the fitting accuracy of the models (AeroMig vs. AeroMG) against measured data using MSE and $R^2$. However, as the objective is not to overfit the data points, as discussed in the previous section, additional metrics such as adjusted $R^2$, AIC, and BIC are incorporated into the composite score metric. This ensures that the proposed AeroMig method maintains strong generalization. The results for MSE, $R^2$, and the composite score metrics are presented in Figs. \ref{fig:MSE_all}, \ref{fig:R2_all}, and \ref{fig:score}, respectively. 

\begin{figure}[htbp]
\centering
\includegraphics[width=0.5\textwidth]{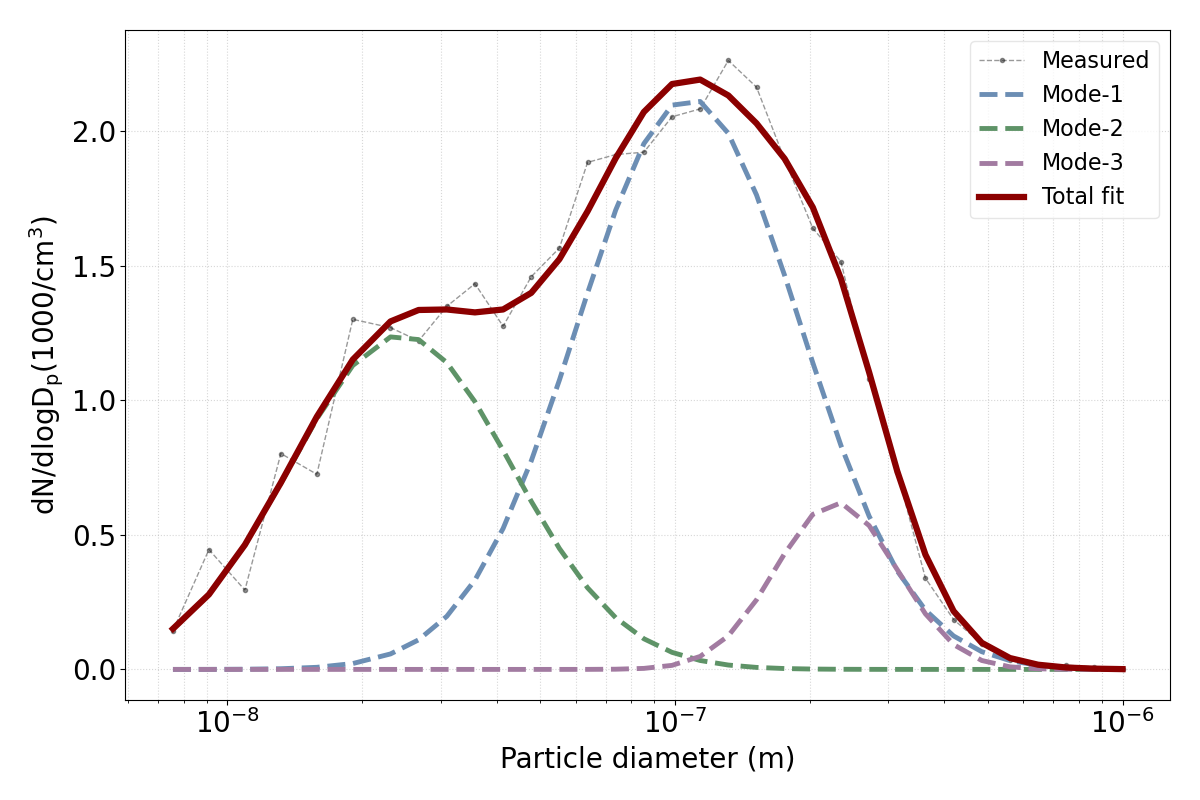}
\vspace{0.5cm}
\includegraphics[width=0.5\textwidth]{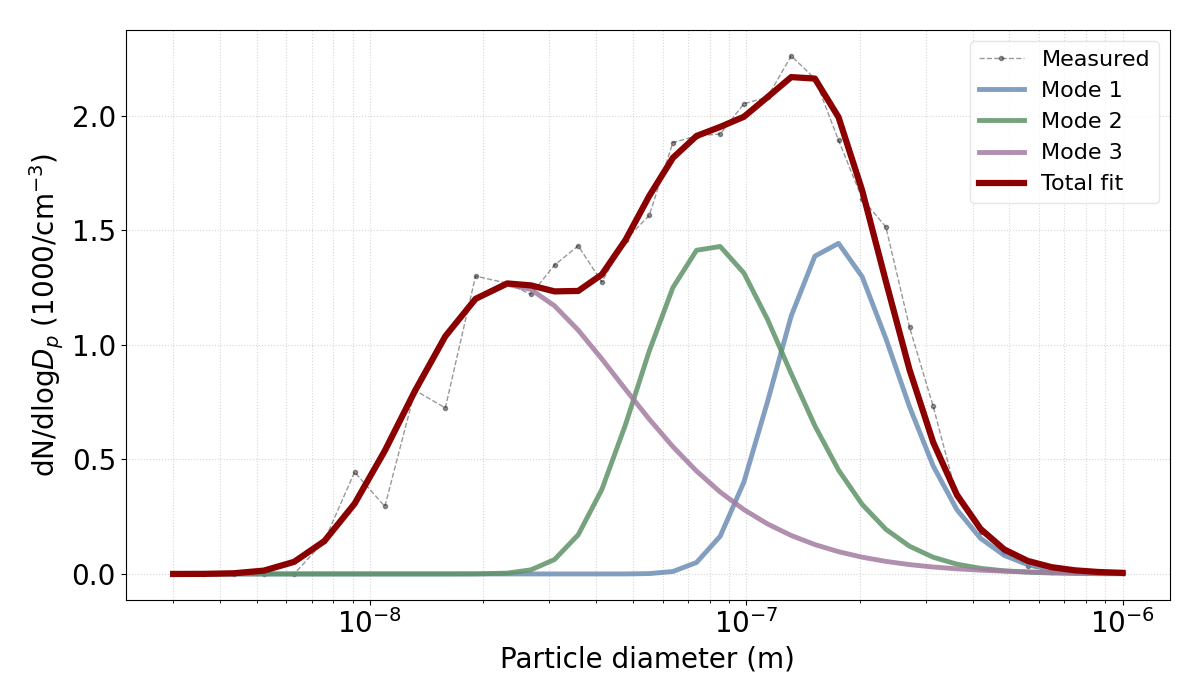}
\caption{Multi-modal fitting profiles for a single continuous PNSD spectrum using the AeroMG (Fig. top 3) and AeroMiG (Fig. bottom 3)}

\label{fig:fitting_row}
\end{figure}

Before evaluating the MSE, $R^2$, and the composite score metrics, we first investigate the underlying modal resolution capabilities of each algorithm by examining representative multi-modal fitting profiles for a single sample continuous spectrum using the AeroMG (Fig. \ref{fig:fitting_row}, top) and AeroMiG (Fig. \ref{fig:fitting_row}, bottom) frameworks. 
Both models decompose the empirical PNSD ($dN/d\log D_p$) into three discrete log-normal sub-modes (Mode $1$, Mode $2$, and Mode $3$) across a particle diameter range from approximately $5\times10 
^{-9}$ \text{m} to $10^{-6}$ \text{m}. 
The AeroMG framework achieves an approximate representation of the global distribution shape, tracking the overall envelope of the measured data points. However, it exhibits localized systemic misfits, notably overestimating the trough near $5\times10^{-8}$ \text{m} and underestimating the maximum peak intensity near $1.5\times10^{-7}$ \text{m}. 
Conversely, the AeroMiG framework demonstrates high structural flexibility, matching both the local fluctuations and the absolute peak amplitudes of the measured data points. The resulting total fit curve closely tracks the experimental data across the entire sub-micron spectrum. This improved continuity is supported by a distinct spatial allocation of the individual sub-modes, where the overlapping boundaries of Modes $1$, $2$, and $3$ closely align with the structural variations present in the raw data points.

We assessed model reliability by plotting the MSE for every individual data point across three different stations (SMEAR I, SMEAR II, and SMEAR III). The fidelity of both methods (AeroMG and AeroMiG) was quantified by calculating the MSE for each fitted PNSD profile. As illustrated in the raincloud plot (Fig. \ref{fig:MSE_all}), AeroMiG demonstrates a lower median MSE and a highly localized error distribution compared to AeroMG. 
The distributional analysis shows that AeroMG exhibits a broad kernel density estimate with a substantial tail of high MSE values. This error tail for AeroMG is most pronounced at the SMEAR III station compared to SMEAR I and SMEAR II. Conversely, the high density of individual data points for AeroMiG remains concentrated at the lower error bound across all three stations.

To evaluate the overall goodness-of-fit across the monitoring stations, we analyzed the coefficient of determination ($R^2$) distribution for both methods (Fig. \ref{fig:R2_all}). At all three sites (SMEAR I, SMEAR II, and SMEAR III), AeroMiG consistently demonstrates superior fitting performance compared to AeroMG, characterized by a higher median $R^2$ value and a narrower, more compressed interquartile range (IQR). 
The raincloud plots reveal a distinct left-skewed distribution for both models, with a high density of individual data points (rain) clustered near $R^2 = 1.0$. However, AeroMG exhibits a substantially longer and denser tail stretching down to the lower bound ($R^2 \approx 0.20$). This performance degradation in AeroMG is progressively exacerbated from SMEAR I to SMEAR III, as evidenced by the increasing density of low-$R^2$ outliers in the scatter components. In contrast, AeroMiG maintains a highly localized upper-bound density across all stations, restricting the majority of its data variance within a tight, high-performing envelope.

To comprehensively evaluate the models across multiple statistical dimensions, we computed a normalized multi-metric score profile encompassing goodness-of-fit ($R^2$, Adjusted $R^2$), scaled information criteria (sAIC, sBIC), and scaled error variance metrics (sMSE) across the three stations (Fig. \ref{fig:score}). The radar charts demonstrate that AeroMiG consistently outperforms AeroMG by enclosing a larger multi-dimensional surface area at all sites. 
While both models achieve nearly identical peak performance along the $R^2$ and Adjusted $R^2$ axes, their trajectories diverge substantially across the remaining parameters. AeroMiG displays a marked expansion along the Score, sAIC, and sBIC axes. Notably, the separation between the two models becomes most pronounced in the sMSE and $R^2$ dimensions. This structural divergence is preserved across the gradient from the clean SMEAR I environment to the highly polluted SMEAR III station, where AeroMG’s performance envelope shrinks further inward along the error-centric axes.

\subsubsection{Computational Efficiency} evaluates the computational time (in seconds) required for each algorithm (AeroMG and AeroMiG) to process a single day of data (consisting of 144 measured data sets). 

\begin{figure}[htbp]
    \centering \includegraphics[width=0.5\textwidth]{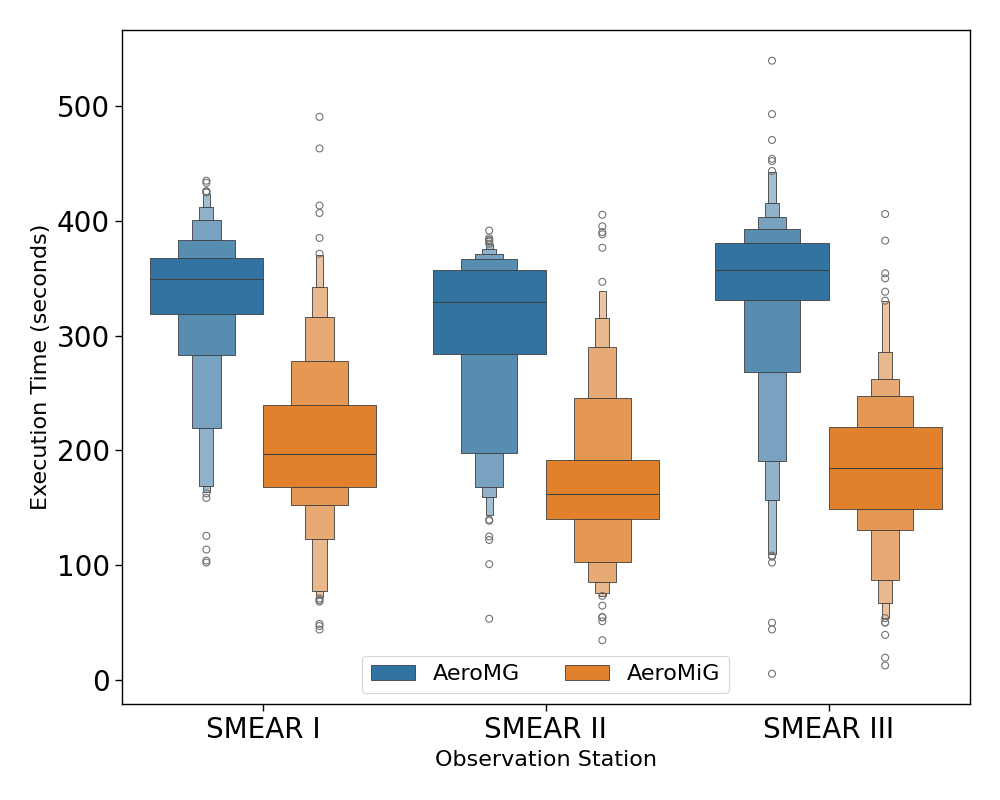}
    \caption{Box-plot of computational time (in seconds) in different stations}
    \label{fig:Time_station_wise}
\end{figure}

To analyze the computational efficiency and operational scalability of both frameworks, we evaluated the execution time distributions (in seconds) across the three tracking stations using letter-value plots (Fig. \ref{fig:Time_station_wise}). AeroMiG consistently demonstrates significantly lower execution times than AeroMG across all sites, reducing the median computational overhead by approximately $40\%$ to $50\%$. At SMEAR I, SMEAR II, and SMEAR III, the median execution times for AeroMiG remain well below $200$ seconds, whereas AeroMG centers around $330$ to $350$ seconds.

Furthermore, the dispersion profile highlights that AeroMiG maintains a narrower and more tightly bounded distribution of computational cost. While both methods display symmetric distributions with a low density of extreme outliers stretching toward higher and lower computation limits, AeroMiG’s upper confidence tiers scale favorably even under more volatile ambient data conditions. This computational advantage persists systematically across the entire station gradient, confirming that the algorithmic speedup is independent of location-specific data features.

\subsection{Discussion}

This subsection discusses the various advantages of the AeroMig algorithm, specifically its consistent performance across different NPF event classifications. It further examines the model's interpretability and assesses its extended capability when extended to handle fitting scenarios involving more than five modes.

\subsubsection{Predictive Accuracies across NPF classes} 
\begin{figure}[htbp]
    \centering    \includegraphics[width=0.5\textwidth]{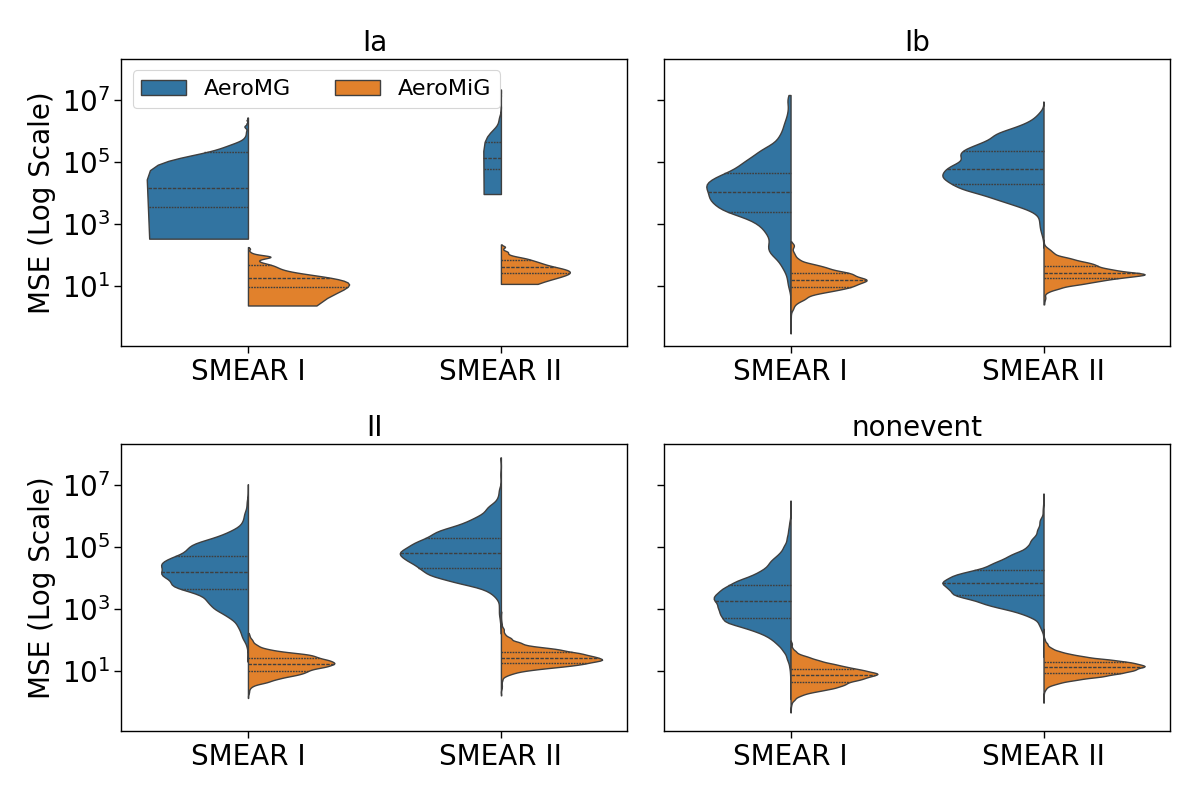}
    \caption{Evaluation of fitting performance across different aerosol event classes. Violin plots showing the distribution of Mean Squared Error (MSE) on a logarithmic scale for AeroMG (blue) and AeroMiG (orange) at the SMEAR I and SMEAR II stations. The distributions are stratified by regional particle formation event classes: Ia, Ib, II, and nonevent. Horizontal dashed lines within the violins indicate the medians and interquartile ranges (IQRs).}
    
    \label{fig:MSE_all_event}
\end{figure}

\begin{figure}[htbp]
    \centering    \includegraphics[width=0.5\textwidth]{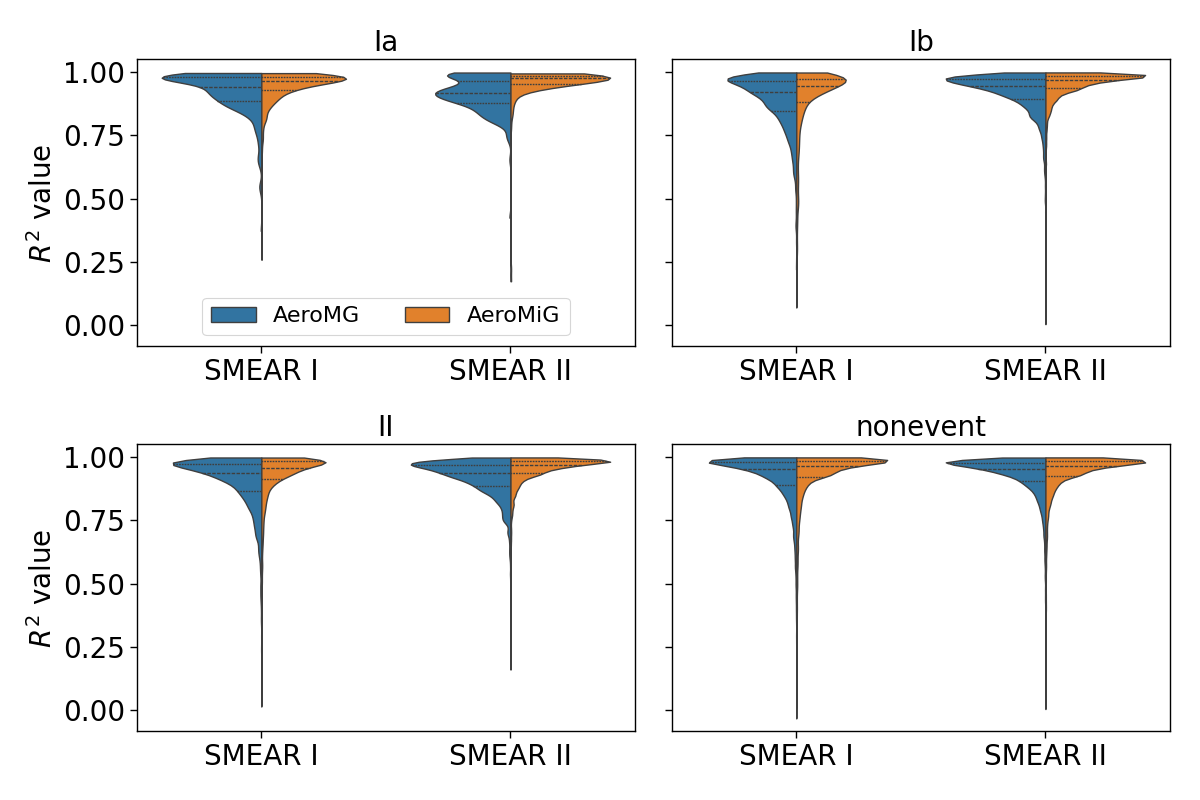}
    \caption{Goodness-of-fit distributions across aerosol event classes. Split-violin plots comparing the coefficient of determination ($R^2$) distributions for AeroMG (blue) and AeroMiG (orange) at the SMEAR I and SMEAR II stations. The data are stratified by atmospheric particle formation event classes (Ia, Ib, II, and nonevent). Horizontal dashed lines indicate the median and interquartile ranges (IQRs) for each model population.}
    \label{fig:r_square_all_event}
\end{figure}

were evaluated to determine if the proposed AeroMig algorithm maintains superior performance compared to AeroMG when both models were fitted to data categorized by NPF event class. Comparisons using MSE and $R^2$ metrics are presented via violin plots in Fig. \ref{fig:MSE_all_event} and Fig. \ref{fig:r_square_all_event}, respectively.
The comparative modal analysis reveals a fundamental divergence in how each framework resolves complex, overlapping aerosol modes. The systemic discrepancies in the AeroMG fit suggest a rigid optimization landscape that often converges on broad, overly smooth sub-modes. This stiffness limits its capacity to handle fine-grained structural fluctuations, leading to the localized over- and under-estimations observed in the accumulation and Aitken mode transition zones.

In contrast, the high fidelity achieved by AeroMiG highlights its superior mathematical adaptability. By allowing more precise, dynamically constrained parameterization of the sub-modes, AeroMiG effectively captures localized peaks and valleys without introducing unphysical mathematical oscillations. The model's ability to closely track the fine variations of the measured distribution confirms that its underlying algorithm successfully balances individual mode constraints with total fit optimization. Consequently, AeroMiG provides a more physically representative decomposition of the aerosol spectrum, which is essential for accurately identifying specific atmospheric processes, such as fresh nucleation tracking, coagulation, and regional transport.

The broad error distribution of AeroMG (Fig. \ref{fig:MSE_all}) suggests that it frequently fails to capture three-mode fitting. This limitation is particularly evident at the SMEAR III station; because SMEAR III features higher pollution and noise levels, AeroMG struggles to accurately fit the noisy data. 
In contrast, AeroMiG's consistently low error variance across diverse aerosol concentrations confirms its superior robustness. Notably, AeroMiG handles polluted data (SMEAR II and III) significantly better than AeroMG. This reduction in error variance is critical for ensuring the overall stability of distribution fitting in highly variable environments.

To verify that the model enhancements are robust against varying atmospheric dynamics, we decoupled the fitting errors by regional aerosol event classification (Fig. \ref{fig:MSE_all_event}). The stratified distributional analysis reveals that AeroMiG consistently delivers superior, low-error performance (MSE $\approx 10^1$) across all event types (Ia, Ib, II, and nonevent) and at both monitoring stations (SMEAR I and SMEAR II).
Conversely, AeroMG shows highly variable performance that degrades depending on the event class. For class Ia events (characterized by highly distinct and clear multi-mode distributions) AeroMG fails completely to resolve lower-error fits, exhibiting truncated distributions bounded at a minimum MSE of approximately $10^2$ to $10^4$. For classes Ib, II, and nonevents, AeroMG exhibits broad, bimodal error distributions with substantial density extending up to an MSE of $10^8$. This widespread variance indicates a frequent, systematic failure to accurately resolve PNSDs, whereas AeroMiG remains tightly bounded and invariant to the specific event conditions.

The comparative $R^2$ analysis (Fig. \ref{fig:R2_all}) underscores the enhanced structural robustness of the AeroMiG framework. While both models achieve high target accuracies under ideal, low-noise conditions (reflected by the shared density peaks near unity), their resilience diverges sharply in challenging environments. The protracted low-$R^2$ tails observed in AeroMG signify recurrent optimization failures or oversimplifications during multi-modal PNSD fitting.

This vulnerability is highly correlated with site environmental characteristics. SMEAR III, typically characterized by elevated background noise and complex multi-modal pollution episodes, induces the highest rate of fitting failures for AeroMG. The fact that AeroMiG suppresses these low-accuracy tails at SMEAR III demonstrates its superior capacity to distinguish true physical aerosol features from complex observational noise. By maintaining structural stability and limiting error variance across diverse ambient concentrations, AeroMiG proves to be a significantly more reliable architecture for automated, long-term atmospheric data inversion.

To complement the error analysis, we evaluated the explanatory power of the models using the coefficient of determination ($R^2$) stratified by event type (Fig. \ref{fig:r_square_all_event}). AeroMiG displays exceptional explanatory fidelity across all classified atmospheric conditions, maintaining a tightly bounded distribution with a median $R^2 > 0.95$ and minimal variance at both SMEAR I and SMEAR II.
In contrast, AeroMG exhibits structural performance degradation and substantial tailing. While its median remains high, the distribution features a prominent downward tail extending toward $R^2 = 0.0$. This degradation is particularly visible in class Ia events at the SMEAR II station, where the lower interquartile bound drops significantly compared to AeroMiG. The highly concentrated, sharp peaks of the AeroMiG violins close to $1.0$ across all event classes demonstrate its superior consistency and statistical robustness in tracking complex, multi-modal aerosol dynamics.

The multi-metric (Fig. \ref{fig:score}) profile highlights that relying solely on conventional goodness-of-fit parameters like $R^2$ can mask underlying model inefficiencies. Although AeroMG matches the maximum fit accuracy of AeroMiG under optimal conditions, its reduced coverage across the scaled information criteria (sAIC, sBIC) and sMSE dimensions reveals a substantial penalty in model parsimony and error management.

The expanded penalization profile of AeroMG suggests a tendency toward overparameterization or optimization instability when resolving complex, multi-modal distributions. Conversely, AeroMiG's superior balance across all axes indicates that its algorithmic architecture achieves a more optimal trade-off between statistical complexity and predictive stability. Because this multi-dimensional superiority is robustly maintained even at the noise-heavy SMEAR III site, the framework demonstrates exceptional structural resilience, making it highly suited for autonomous, high-fidelity atmospheric data processing where data quality varies drastically.

The substantial reduction in execution time (Fig. \ref{fig:Time_station_wise}) achieved by AeroMiG underscores its algorithmic optimization and suitability for high-throughput atmospheric applications. In large-scale climate monitoring and observational networks, processing multi-decadal PNSD data manually or with computationally heavy inversion techniques creates severe processing bottlenecks. By halving the computational footprint while simultaneously achieving superior fitting accuracy and structural stability (as demonstrated in the preceding $R^2$ and multi-metric analyses), AeroMiG breaks the traditional trade-off between statistical fidelity and computational cost.

This enhanced speed can likely be attributed to a more efficient convergence pathway within the AeroMiG optimization framework, preventing the algorithm from getting trapped in computationally expensive local minima or redundant iterations. Because the model limits execution times to a predictable and compressed envelope across different data types, it presents a highly scalable architecture. This computational efficiency is vital for enabling real-time or near-real-time automated data processing pipelines across global aerosol monitoring networks.

The close agreement between the data before fitting and after fitting demonstrates the structural fidelity of the numerical framework during dynamic atmospheric phenomena. Resolving NPF event types poses a substantial challenge for automated fitting routines because of the rapid, orders-of-magnitude changes in concentration and the shifting boundaries between nucleation, Aitken, and accumulation modes. The framework’s ability to replicate the distinct asymmetry of Class Ia events without over-smoothing the peak concentration indicates an exceptional capability to handle highly localized gradients.

Furthermore, preserving the precise variance envelopes ($1\sigma$) after fitting has major implications for downstream statistical and climate modeling. The high concentration variance observed in the nucleation mode during Class Ia episodes reflects the burst-like, discrete nature of regional nucleation and subsequent growth. By accurately capturing both the mean state and the underlying physical variance of the data (while separating them from the stable, accumulation-dominated background of nonevents), the model proves it can act as a reliable surrogate for automated event classification pipelines. This high-fidelity reproduction of the raw data trends and structural noise boundaries confirms the model's readiness for implementation in long-term, continuous global climate monitoring arrays.

\begin{figure}[htbp]
\centering
\includegraphics[width=0.5\textwidth]{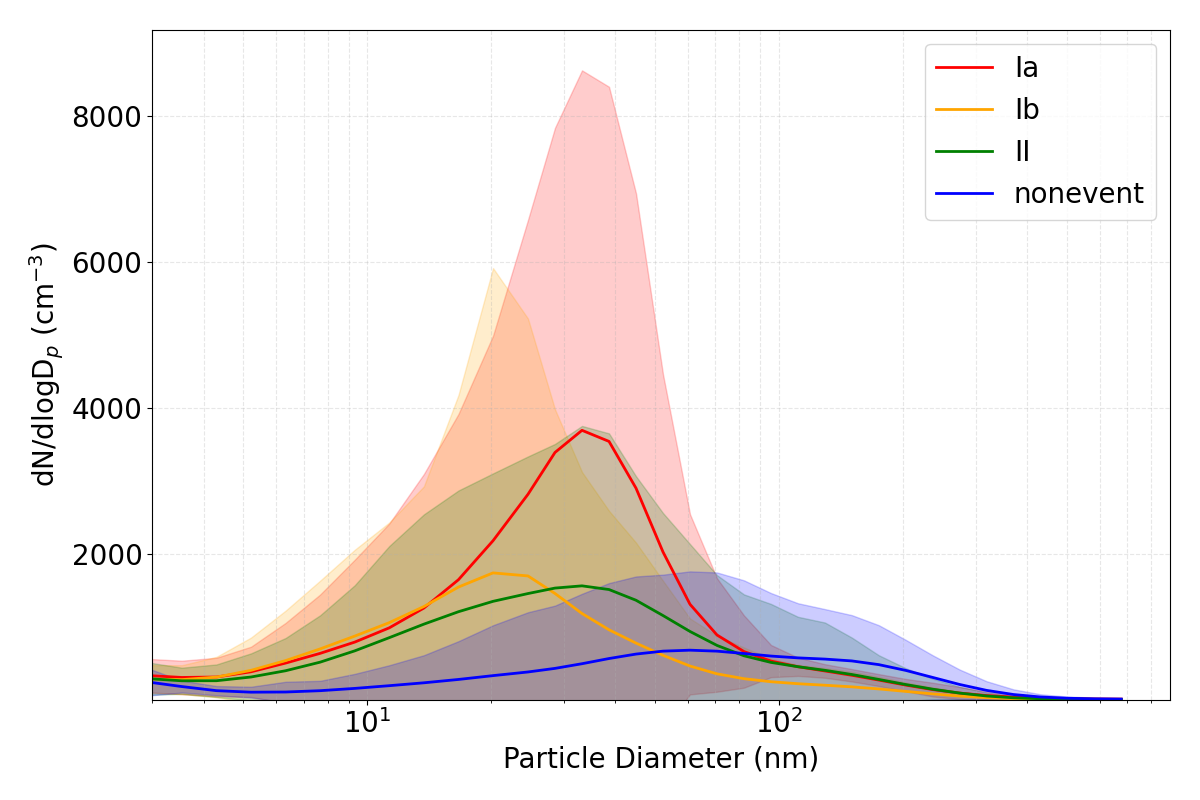}
\vspace{0.5cm}
\includegraphics[width=0.5\textwidth]{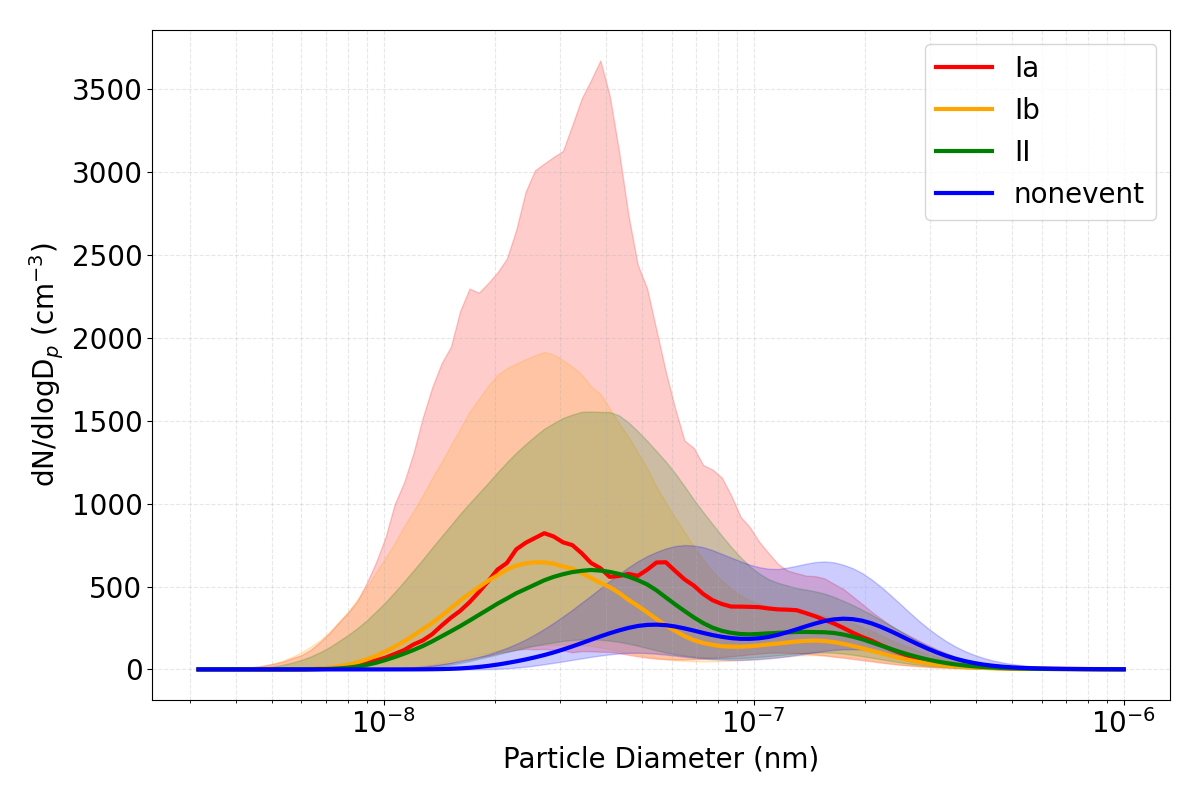}
\caption{Multi-modal Inverse-Gamma parameterization of aerosol size distributions across NPF classes. Top, Observed sub-micrometer PNSDs ($dN/d\log D_p$) at the SMEAR II station, categorized by NPF event strength (Classes Ia, Ib, II) and nonevent days (solid lines indicate means; shaded regions indicate $\pm1\sigma$). Down, Reconstructed distributions parameterized via multi-modal Inverse-Gamma functions. The mathematical transition from strong nucleation (Class Ia) to nonevent conditions is driven by a systematic reduction in the shape parameter ($\hat \alpha$) and an increase in the scale parameter ($\hat \beta$) across the distribution modes (Table \ref{tab:parameter}). High $\hat \alpha$ values in Class Ia produce the sharp, tightly bounded peaks characteristic of highly coordinated regional nucleation. Conversely, the suppressed $\hat \alpha$ and elevated $\hat \beta$ values during nonevents broaden the distribution, mathematically capturing the transition from deterministic particle growth tracks to a highly dispersed, stochastic background aerosol regime.}

\label{fig:concentration_class_fit}
\end{figure}

\subsubsection{Interpretability} is a key feature that ensures AeroMig remains transparent and accessible for analysis. While AeroMG is based on the log-normal (log-Gaussian) distribution, which is inherently easy to interpret as described in the previous section, the parameters of AeroMig can also be mapped directly to mean and standard deviation values. As a case study, we aggregated all PNSD data from the SMEAR I station according to NPF classification, as shown in the upper panel of Fig. \ref{fig:concentration_class_fit}. We then applied the AeroMig algorithm to this aggregated data, with the results displayed in both subfigures of Fig. \ref{fig:concentration_class_fit}. The results demonstrate that the fitted curves provide a more structured representation of the physical properties inherent in PNSD data. Furthermore, we extracted the mean and standard deviation variables from the $\alpha$ and $\beta$ parameters to demonstrate that every mode in AeroMig is as interpretable as those in AeroMG. 

To translate the model's latent feature space into physically interpretable physical states, we mapped the learned representations to Inverse-Gamma distributions across distinct NPF event classes (Table \ref{tab:parameter}). While the estimated shape ($\hat\alpha$) and scale ($\hat\beta$) parameters define the mathematical topology of the distribution, we explicitly converted these parameters into the first two statistical moments (mean and standard deviation (std)) to provide a direct proxy for physical atmospheric dynamics.
This conversion uncovers a clear parametric gradient that tracks NPF event strength and clarity:
Class Ia and Ib (strong nucleation) are characterized by the lowest mean values (e.g., mean$_1 = 1.17\times 10^{-7}$ for Ia) and tightly bounded standard deviations (std$_1 = 5.50\times 10^{-8}$). Physically, these low-mean, low-variance profiles represent highly constrained atmospheric conditions. They signify a deterministic and uniform physical trigger, such as a sharp influx of precursor vapors paired with a low pre-existing condensation sink, which is characteristic of ideal "banana-shaped" growth tracks at the SMEAR II station.

As the NPF signatures (class II to nonevents, weak/no nucleation) degrade into nonevent days, the mean and standard deviation systematically increase across all three distribution modes. For instance, the first mode's mean more than doubles ($2.64\times 10^{-7}$), and its standard deviation nearly doubles ($9.36\times 10^{-8}$) on nonevent days.
Physically, this expansion of the mean and variance captures the transition from highly synchronized aerosol nucleation to a stochastic, unconstrained atmospheric regime. The broader standard deviation on nonevent days directly quantifies the increased environmental noise, background aerosol fluctuations, and regional transport effects that suppress steady nucleation. By converting abstract algebraic parameters into tangible physical moments (mean and std), the model's internal representations become fully transparent, acting as a quantifiable diagnostic tool for atmospheric stability and particle distribution behavior.

To assess the capacity of the framework to resolve physical atmospheric features across diverse conditions, we analyzed the PNSDs ($dN/d\log D_p$) and associated variances ($1\sigma$ shaded regions) before fitting (Fig. top \ref{fig:concentration_class_fit}) and after fitting (Fig. bottom \ref{fig:concentration_class_fit}), stratified by NPF event types (Class Ia, Ib, II, and nonevent).

\begin{table}[ht]
\centering
\footnotesize
\caption{Inverse-Gamma Parameters and Distribution Statistics (SMEAR II)}
\begin{tabular}{lcccc}
\toprule
Metric & Ia & Ib & II & nonevent \\
\midrule
$\hat\alpha_1$ & $8.23\times 10^{+0}$ & $7.71\times 10^{+0}$ & $7.60\times 10^{+0}$ & $7.22\times 10^{+0}$ \\
$\hat\beta_1$  & $8.78\times 10^{-7}$ & $8.76\times 10^{-7}$ & $9.10\times 10^{-7}$ & $1.20\times 10^{-6}$ \\
mean$_1$       & $1.17\times 10^{-7}$ & $1.85\times 10^{-7}$ & $2.25\times 10^{-7}$ & $2.64\times 10^{-7}$ \\
std$_1$        & $5.50\times 10^{-8}$ & $6.15\times 10^{-8}$ & $6.45\times 10^{-8}$ & $9.36\times 10^{-8}$ \\
\midrule
$\hat\alpha_2$ & $8.52\times 10^{+0}$ & $7.70\times 10^{+0}$ & $7.68\times 10^{+0}$ & $6.93\times 10^{+0}$ \\
$\hat\beta_2$  & $8.81\times 10^{-7}$ & $8.77\times 10^{-7}$ & $9.31\times 10^{-7}$ & $1.09\times 10^{-6}$ \\
mean$_2$       & $1.37\times 10^{-7}$ & $1.74\times 10^{-7}$ & $2.09\times 10^{-7}$ & $4.32\times 10^{-7}$ \\
std$_2$        & $5.19\times 10^{-8}$ & $6.08\times 10^{-8}$ & $6.61\times 10^{-8}$ & $9.40\times 10^{-8}$ \\
\midrule
$\hat\alpha_3$ & $8.28\times 10^{+0}$ & $7.74\times 10^{+0}$ & $7.67\times 10^{+0}$ & $7.20\times 10^{+0}$ \\
$\hat\beta_3$  & $9.80\times 10^{-7}$ & $9.58\times 10^{-7}$ & $1.00\times 10^{-6}$ & $1.23\times 10^{-6}$ \\
mean$_3$       & $1.53\times 10^{-7}$ & $1.72\times 10^{-7}$ & $2.96\times 10^{-7}$ & $2.92\times 10^{-7}$ \\
std$_3$        & $6.95\times 10^{-8}$ & $6.22\times 10^{-8}$ & $6.47\times 10^{-8}$ & $9.46\times 10^{-8}$ \\
\bottomrule
\end{tabular}
\label{tab:parameter}
\end{table}

The profiles before fitting reveal distinct, event-specific multi-modal structures. Class Ia events are characterized by a sharp, dominant nucleation/Aitken mode peaking near $30\text{-}40$ \text{nm} with a maximum concentration exceeding $3500$ \text{cm}$^{-3}$, accompanied by a broad, highly volatile envelope extending up to $8500$ \text{cm}$^{-3}$. Class Ib and II events present progressively broader, suppressed nucleation peaks shifting toward slightly larger diameters ($20\text{-}50$ \text{nm}) and lower mean amplitudes. Conversely, the nonevent profiles exhibit a characteristic accumulation-mode bias, peaking weakly between $100$ and $200$ \text{nm} ($\sim 500$ \text{cm}$^{-3})$ with minimal particle density under $20$ \text{nm}.

A direct comparison with the profiles after fitting confirms that the framework replicates these highly non-linear, class-specific modal shapes and boundaries with high fidelity. The mathematical inversion accurately captures the steep concentration gradients and maps the shifting peak diameters across all event classes. Crucially, the $1\sigma$ variance bounds after fitting closely mirror the structural geometry of the data before fitting, capturing the high variability of active nucleation bursts (Class Ia) while preserving the tighter, more stable envelopes seen in nonevent regimes.

\begin{figure}[htbp]
\centering
\includegraphics[width=0.5\textwidth]{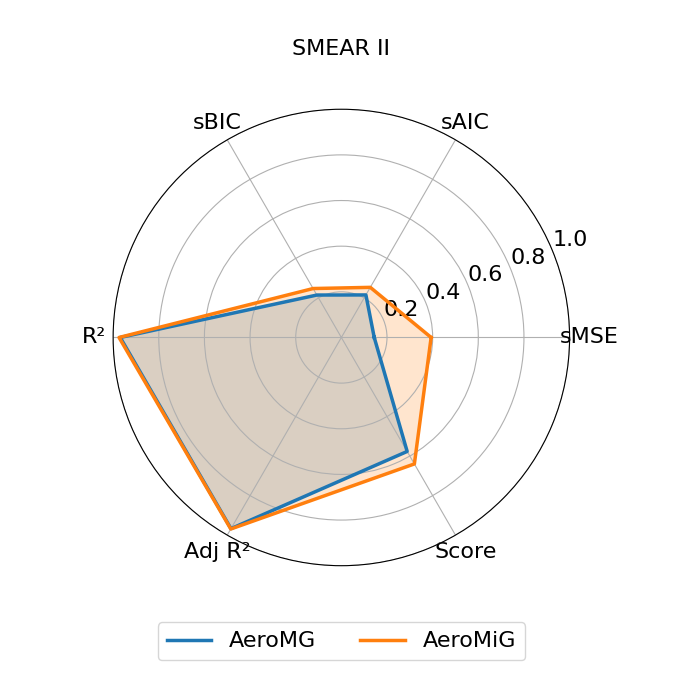}
\vspace{0.5cm}
\includegraphics[width=0.5\textwidth]{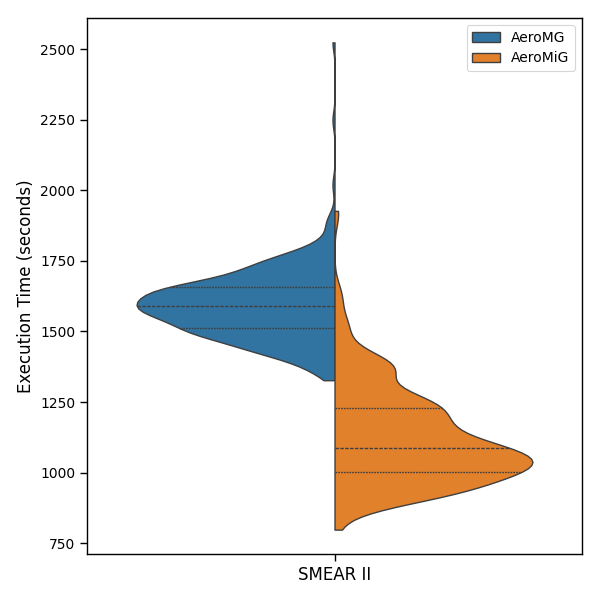}
\caption{Multi-criteria performance trade-offs and computational efficiency for 5-mode fitting. Top, Radar chart evaluating overall model fit at the SMEAR II station across six standardized statistical metrics: scaled Bayesian Information Criterion (sBIC), scaled Akaike Information Criterion (sAIC), scaled Mean Squared Error (sMSE), holistic composite Score, Adjusted $R^2$, and $R^2$. Larger shaded areas indicate superior multi-objective performance. Down, Split-violin plot comparing the total execution time (seconds) distributions for AeroMG (blue) and AeroMiG (orange) optimization runs. Horizontal dashed lines denote medians and interquartile ranges (IQRs).}

\label{fig:fitting_5_mode}
\end{figure}

\subsubsection{Extended Capability} evaluate the AeroMig algorithm under realistic field conditions, we utilized a multi-instrument field dataset collected at the SMEAR II station spanning a continuous year period from May 2019 to April 2020 (\cite{sulo_long-term_2021}). The preprocessing framework dynamically unifies raw measurements from a Particle Size Magnifier (PSM), a Differential Mobility Particle Sizer (DMPS), and an Aerodynamic Particle Sizer (APS) into a single master time-series matrix. To resolve the sub\text{-}$3$ \text{nm} nucleation track, raw counts from the PSMs $1.1$\text{-}$1.7$ \text{nm} tracking channel are normalized to a differential concentration $dN/d\log D_p$) and assigned to a representative geometric mean diameter ($D_g$) of $1.366$ \text{nm}. For the upper coarse boundary, aerodynamic diameters from the APS are iteratively converted to mobility equivalents using the Cunningham slip correction factor assuming an effective particle density of $1.6$ \text{g/cm}$^3$ and a dynamic shape factor ($\chi$) of $1.0$, terminating at a maximum diameter of $15,074$ \text{nm} ($1.5073558\times10^{-5}$ \text{m}). The resulting synchronized data array forms the basis for evaluating the five-mode algorithm configuration.

To evaluate the multi-objective trade-offs between fitting accuracy, model parsimony, and computational efficiency, we conducted a comprehensive evaluation at the SMEAR II station (Fig. \ref{fig:fitting_5_mode}). The radar profile (Fig. \ref{fig:fitting_5_mode} top) demonstrates that AeroMiG outperforms or matches AeroMG across all six standardized statistical criteria. While both frameworks exhibit comparable raw explanatory power (reaching maximum values near $1.0$ for $R^2$ and Adjusted $R^2$), AeroMiG achieves a markedly larger multi-objective area. This superiority is driven by its substantially higher composite performance Score and optimized information criteria (sAIC, sBIC, and sMSE), proving that its structural design better balances optimization depth with statistical penalization.
Crucially, this expanded performance profile does not come at a computational premium. As shown by the execution time distributions (Fig. \ref{fig:fitting_5_mode} down), AeroMiG provides an approximate $35\%$ reduction in median processing time compared to AeroMG. The execution times for AeroMiG are tightly bounded, with a median of approximately $1000$ \text{s} and an absence of high-latency tails. Conversely, AeroMG exhibits a higher median execution time ($\sim 1550$ \text{s}) alongside a protracted high-latency tail stretching past $2500$ \text{s}. This combined efficiency and optimization stability establishes AeroMiG as a highly scalable framework for 5-mode large-scale, high-throughput atmospheric data inversion.

\subsection{Practical uses of the proposed method}

The proposed AeroMiG framework enables rapid, stable characterization of PNSDs with efficient convergence over highly skewed structures, facilitating its seamless integration into real-time air quality systems for the immediate detection of sharp aerosol variations driven by pollution events like traffic emissions or industrial spikes. Because different aerosol sources emit highly distinct particle size profiles, the compact modal descriptors extracted by AeroMiG can be utilized in downstream machine learning pipelines to automate aerosol classification and source attribution. This approach successfully addresses key challenges in urban environmental monitoring, such as differentiating between local, fresh vehicular emissions and regional, long-range-transported backgrounds solely from size distribution dynamics \cite{harni_source_2024}. Furthermore, by mapping out precise mathematical descriptors across varying locations and timeframes, the framework helps resolve broader global environmental challenges, including the historical difficulty of characterizing highly mismatched particle number concentrations versus mass concentrations across diverse global microenvironments \cite{de_jesus_ultrafine_2019}.

\section{Conclusion}

This study addresses a central challenge in atmospheric data analysis such that the accurate representation of PNSDs, which are inherently skewed, heavy-tailed, and multi-modal. We introduced a flexible parametric framework, AeroMiG, that models these distributions as mixtures of Inverse-Gamma components, providing a more faithful representation of aerosol size structure than conventional lognormal-based approaches.
Evaluation using real-world measurements from three independent monitoring stations demonstrates that the proposed method consistently improves fitting accuracy, reduces error, and enhances computational efficiency relative to the existing technique. These improvements are reflected across multiple performance metrics, including $\mathrm{R}^2$, adjusted $\mathrm{R}^2$, MSE, AIC, BIC, and a composite performance score, indicating both statistical robustness and practical reliability.
Beyond performance gains, the proposed framework offers a conceptually meaningful representation of aerosol populations by capturing asymmetry and tail behavior in a physically interpretable manner. This enables more accurate characterization of aerosol processes and provides a scalable foundation for analyzing large observational datasets.

The broader significance of this work lies in its potential to support real-time air quality monitoring, improve data-driven atmospheric modeling, and enhance the integration of observational data into predictive systems. More generally, the framework provides a transferable approach for modeling complex, heavy-tailed distributions across diverse scientific and engineering domains.
Future work will focus on adaptive strategies for selecting the optimal number of mixture components and extending the framework to dynamic and high-dimensional environmental datasets.

\section*{Acknowledgements}

This project is supported by the Research Council of Finland
through the Academy Research Fellowship projects (Grant
No. 355330 and 364223) and the University Profiling funding initiative InterEarth (Grant No. 353218), Center of Excellence in Atmospheric Sciences (ACCS) (Grant No. 356134) and PROFI for the University of Eastern Finland (Grant No. 325022 and 352968).

We acknowledge the use of OpenAI’s ChatGPT-4 (https://chat.openai.com/) and Google's Gemini-3.5 Flash (https://gemini.google.com/) to refine the academic language and coherence of selected sections.

\bibliographystyle{IEEEtran}
\bibliography{references.bib}

@article{li_scattering_2022,
	title = {Scattering and absorbing aerosols in the climate system},
	volume = {3},
	issn = {2662-138X},
	url = {https://www.nature.com/articles/s43017-022-00296-7},
	doi = {10.1038/s43017-022-00296-7},
	pages = {363--379},
	number = {6},
	journal = {Nature Reviews Earth \& Environment},
	shortjournal = {Nat Rev Earth Environ},
	author = {Li, Jing and Carlson, Barbara E. and Yung, Yuk L. and Lv, Daren and Hansen, James and Penner, Joyce E. and Liao, Hong and Ramaswamy, V. and Kahn, Ralph A. and Zhang, Peng and Dubovik, Oleg and Ding, Aijun and Lacis, Andrew A. and Zhang, Lu and Dong, Yueming},
	urldate = {2025-11-26},
    year = {2022},
	date = {2022-05-24},
	langid = {english},
}

@article{pope_iii_lung_2002,
	title = {Lung Cancer, Cardiopulmonary Mortality, and Long-term Exposure to Fine Particulate Air Pollution},
	volume = {287},
	issn = {0098-7484},
	url = {http://jama.jamanetwork.com/article.aspx?doi=10.1001/jama.287.9.1132},
	doi = {10.1001/jama.287.9.1132},
	pages = {1132},
	number = {9},
	journal = {{JAMA}},
	shortjournal = {{JAMA}},
	author = {Pope Iii, C. Arden},
	urldate = {2026-03-28},
    year = {2002},
	date = {2002-03-06},
	langid = {english},
}

@book{stocker_climate_2014,
	location = {New York},
	title = {Climate change 2013: the physical science basis: Working Group I contribution to the Fifth assessment report of the Intergovernmental Panel on Climate Change},
	isbn = {978-1-107-05799-9 978-1-107-66182-0},
	shorttitle = {Climate change 2013},
	pagetotal = {1535},
	publisher = {Cambridge University Press},
	editor = {Stocker, Thomas},
    year = {2014},
	date = {2014},
}

@article{masson2021climate,
  title={Climate change 2021: the physical science basis},
  author={Masson-Delmotte, Val{\'e}rie and Zhai, Panmao and Pirani, Anna and Connors, Sarah L and P{\'e}an, Clotilde and Berger, Sophie and Caud, Nada and Chen, Y and Goldfarb, L and Gomis, MI and others},
  journal={Contribution of working group I to the sixth assessment report of the intergovernmental panel on climate change},
  volume={2},
  number={1},
  pages={2391},
  year={2021},
  publisher={Cambridge University Press Cambridge, UK}
}

@book{intergovernmental_panel_on_climate_change_ipcc_climate_2023,
	edition = {1},
	title = {Climate Change 2021 – The Physical Science Basis: Working Group I Contribution to the Sixth Assessment Report of the Intergovernmental Panel on Climate Change},
	rights = {https://www.cambridge.org/core/terms},
	isbn = {978-1-009-15789-6},
	url = {https://www.cambridge.org/core/product/identifier/9781009157896/type/book},
	shorttitle = {Climate Change 2021 – The Physical Science Basis},
	publisher = {Cambridge University Press},
	author = {{Intergovernmental Panel On Climate Change (Ipcc)}},
	urldate = {2025-11-26},
    year = {2023},
	date = {2023-07-06},
	doi = {10.1017/9781009157896},
}

@article{whitby_physical_1978,
	title = {The physical characteristics of sulfur aerosols},
	volume = {12},
	rights = {https://www.elsevier.com/tdm/userlicense/1.0/},
	issn = {00046981},
	url = {https://linkinghub.elsevier.com/retrieve/pii/0004698178901968},
	doi = {10.1016/0004-6981(78)90196-8},
	pages = {135--159},
	number = {1},
	journal = {Atmospheric Environment (1967)},
	shortjournal = {Atmospheric Environment (1967)},
	author = {Whitby, Kenneth T.},
	urldate = {2026-03-28},
	date = {1978-01},
    year = {1978},
	langid = {english},
}

@article{akinyoola_review_2024,
	title = {A Review of Atmospheric Aerosol Impacts on Regional Extreme Weather and Climate Events},
	volume = {8},
	issn = {2510-375X, 2510-3768},
	url = {https://link.springer.com/10.1007/s41810-024-00223-x},
	doi = {10.1007/s41810-024-00223-x},
	pages = {249--274},
	number = {3},
	journal = {Aerosol Science and Engineering},
	shortjournal = {Aerosol Sci Eng},
	author = {Akinyoola, Julius A. and Oluleye, A. and Gbode, Imoleayo E.},
	urldate = {2026-03-28},
    year = {2024},
	date = {2024-09},
	langid = {english},
}

@article{storn_differential_1997,
	title = {Differential Evolution – A Simple and Efficient Heuristic for global Optimization over Continuous Spaces},
	volume = {11},
	rights = {https://www.springernature.com/gp/researchers/text-and-data-mining},
	issn = {0925-5001, 1573-2916},
	url = {https://link.springer.com/10.1023/A:1008202821328},
	doi = {10.1023/A:1008202821328},
	pages = {341--359},
	number = {4},
	journal = {Journal of Global Optimization},
	shortjournal = {Journal of Global Optimization},
	author = {Storn, Rainer and Price, Kenneth},
	urldate = {2026-02-09},
    year = {1997},
	date = {1997-12},
	langid = {english},
}

@article{dal2005formation,
  title={Formation and growth of fresh atmospheric aerosols: eight years of aerosol size distribution data from SMEAR II, Hyyti{\"a}l{\"a}, Finland},
  author={Dal Maso, Miikka and Kulmala, Markku and Riipinen, Ilona and Wagner, Robert and Hussein, Tareq and Aalto, Pasi P and Lehtinen, Kari EJ},
  journal={Boreal environment research},
  volume={10},
  number={5},
  pages={323},
  year={2005},
  publisher={Finnish Environment Institute}
}

@article{curtius_nucleation_2006,
	title = {Nucleation of atmospheric aerosol particles},
	volume = {7},
	issn = {1631-0705, 1878-1535},
	url = {https://comptes-rendus.academie-sciences.fr/physique/articles/10.1016/j.crhy.2006.10.018/},
	doi = {10.1016/j.crhy.2006.10.018},
	pages = {1027--1045},
	number = {9},
	journal = {Comptes Rendus. Physique},
	author = {Curtius, Joachim},
	urldate = {2025-11-25},
    year = {2006},
	date = {2006-11-01},
	langid = {english},
}

@article{whitby_modal_1997,
	title = {Modal Aerosol Dynamics Modeling},
	volume = {27},
	issn = {0278-6826, 1521-7388},
	url = {http://www.tandfonline.com/doi/abs/10.1080/02786829708965504},
	doi = {10.1080/02786829708965504},
	pages = {673--688},
	number = {6},
	journal = {Aerosol Science and Technology},
	shortjournal = {Aerosol Science and Technology},
	author = {Whitby, Evan R. and {McMurry}, Peter H.},
	urldate = {2025-11-25},
    year = {1997},
	date = {1997-01},
	langid = {english},
}

@book{seinfeld_atmospheric_2016,
	location = {Newark},
	edition = {1st ed},
	title = {Atmospheric Chemistry and Physics: From Air Pollution to Climate Change},
	isbn = {978-1-118-94740-1 978-1-119-22116-6},
	series = {New York Academy of Sciences Series},
	shorttitle = {Atmospheric Chemistry and Physics},
	pagetotal = {1},
	publisher = {John Wiley \& Sons, Incorporated},
	author = {Seinfeld, John H.},
    year = {2016},
	date = {2016},
}

@article{zhu_comprehensive_2024,
	title = {A comprehensive study on the validation and application of multi-lognormal distribution models for atmospheric particles},
	volume = {338},
	issn = {13522310},
	url = {https://linkinghub.elsevier.com/retrieve/pii/S1352231024004886},
	doi = {10.1016/j.atmosenv.2024.120813},
	pages = {120813},
	journal = {Atmospheric Environment},
	shortjournal = {Atmospheric Environment},
	author = {Zhu, Ke and Wang, Lina},
	urldate = {2026-03-29},
    year = {2024},
	date = {2024-12},
	langid = {english}
}

@article{Hussein2005,
  title={Evaluation of an automatic algorithm for fitting the particle number size distributions},
  author={Tareq Hussein and Miikka Dal Maso and Tuukka Pet{\"a}j{\"a} and Ismo Kalevi Koponen and Pentti Paatero and Pasi P. Aalto and Kaarle J. H{\"a}meri and Markku Kulmala},
  journal={Boreal Environment Research},
  year={2005},
  volume={10},
  pages={337-355},
  url={https://api.semanticscholar.org/CorpusID:73658875}
}

@book{hinds_aerosol_2022,
	location = {Hoboken, {NJ}},
	title = {Aerosol technology: properties, behavior, and measurement of airborne particles},
	isbn = {978-1-119-49404-1},
	shorttitle = {Aerosol technology},
	pagetotal = {425},
	publisher = {Wiley},
	author = {Hinds, William C. and Zhu, Yifang},
    year = {2022},
	date = {2022}
}

@book{colbeck_aerosol_2014,
	location = {West Sussex, United Kingdom},
	title = {Aerosol science: technology and applications},
	isbn = {978-1-119-97792-6 978-1-118-67535-9},
	shorttitle = {Aerosol science},
	pagetotal = {1},
	publisher = {John Wiley \& Sons},
	editor = {Colbeck, Ian and Lazaridis, Mihalis},
    year = {2014},
	date = {2014},
}

@book{bishop_pattern_2016,
	location = {New York, {NY}},
	note = {Softcover reprint of the original edition 2006 (corrected at 8th printing 2009)},
	title = {Pattern Recognition and Machine Learning},
	isbn = {978-1-4939-3843-8},
	series = {Information science and statistics},
	pagetotal = {738},
	publisher = {Springer New York},
	author = {Bishop, Christopher M.},
    year = {2016},
	date = {2016},
}

@book{huber_robust_2009,
	edition = {1},
	title = {Robust Statistics},
	rights = {http://doi.wiley.com/10.1002/tdm\_license\_1.1},
	isbn = {978-0-470-12990-6 978-0-470-43469-7},
	url = {https://onlinelibrary.wiley.com/doi/book/10.1002/9780470434697},
	series = {Wiley Series in Probability and Statistics},
	publisher = {Wiley},
	author = {Huber, Peter J. and Ronchetti, Elvezio M.},
	urldate = {2025-10-16},
    year = {2009},
	date = {2009-01-29},
	langid = {english},
	doi = {10.1002/9780470434697},
}

@book{murphy_machine_2012,
	location = {Cambridge, Mass.},
	title = {Machine learning: a probabilistic perspective},
	isbn = {978-0-262-30524-2},
	shorttitle = {Machine learning},
	publisher = {{MIT} Press},
	author = {Murphy, Kevin P.},
    year = {2012},
	date = {2012},
	note = {{OCLC}: 810414751},
}

@article{zhao_infovae_2019,
	title = {{InfoVAE}: Balancing Learning and Inference in Variational Autoencoders},
	volume = {33},
	rights = {https://www.aaai.org},
	issn = {2374-3468, 2159-5399},
	url = {https://ojs.aaai.org/index.php/AAAI/article/view/4538},
	doi = {10.1609/aaai.v33i01.33015885},
	shorttitle = {{InfoVAE}},
	pages = {5885--5892},
	number = {1},
	journal = {Proceedings of the {AAAI} Conference on Artificial Intelligence},
	shortjournal = {{AAAI}},
	author = {Zhao, Shengjia and Song, Jiaming and Ermon, Stefano},
	urldate = {2025-10-16},
    year = {2019},
	date = {2019-07-17},
}

@article{nyaku_comparison_2020,
	title = {A comparison of lognormal and gamma size distributions for characterizing the stratospheric aerosol phase function from optical particle counter measurements},
	volume = {13},
	rights = {https://creativecommons.org/licenses/by/4.0/},
	issn = {1867-8548},
	url = {https://amt.copernicus.org/articles/13/1071/2020/},
	doi = {10.5194/amt-13-1071-2020},
	pages = {1071--1087},
	number = {3},
	journal = {Atmospheric Measurement Techniques},
	shortjournal = {Atmos. Meas. Tech.},
	author = {Nyaku, Ernest and Loughman, Robert and Bhartia, Pawan K. and Deshler, Terry and Chen, Zhong and Colarco, Peter R.},
	urldate = {2025-10-16},
    year = {2020},
	date = {2020-03-06},
	langid = {english},
}

@article{ren_prediction_2020,
	title = {Prediction of Aerosol Particle Size Distribution Based on Neural Network},
	volume = {2020},
	rights = {http://creativecommons.org/licenses/by/4.0/},
	issn = {1687-9309, 1687-9317},
	url = {https://www.hindawi.com/journals/amete/2020/5074192/},
	doi = {10.1155/2020/5074192},
	pages = {1--11},
	journal = {Advances in Meteorology},
	shortjournal = {Advances in Meteorology},
	author = {Ren, Yali and Mao, Jiandong and Zhao, Hu and Zhou, Chunyan and Gong, Xin and Rao, Zhimin and Wang, Qiang and Zhang, Yi},
	urldate = {2025-10-20},
    year = {2020},
	date = {2020-06-06},
	langid = {english},
}

@book{hastie_elements_2009,
	location = {New York, {NY}},
	title = {The Elements of Statistical Learning},
	rights = {http://www.springer.com/tdm},
	isbn = {978-0-387-84857-0 978-0-387-84858-7},
	url = {http://link.springer.com/10.1007/978-0-387-84858-7},
	series = {Springer Series in Statistics},
	publisher = {Springer New York},
	author = {Hastie, Trevor and Tibshirani, Robert and Friedman, Jerome},
	urldate = {2025-10-20},
    year = {2009},
	date = {2009},
	doi = {10.1007/978-0-387-84858-7},
}

@article{sugiura_further_1978,
	title = {Further analysis of the data by Akaike's information criterion and the finite corrections: Further analysis of the data by akaike' s},
	volume = {7},
	issn = {0361-0926, 1532-415X},
	url = {https://www.tandfonline.com/doi/full/10.1080/03610927808827599},
	doi = {10.1080/03610927808827599},
	shorttitle = {Further analysis of the data by Akaike's information criterion and the finite corrections},
	pages = {13--26},
	number = {1},
	journal = {Communications in Statistics - Theory and Methods},
	shortjournal = {Communications in Statistics - Theory and Methods},
	author = {Sugiura, Nariaki},
	urldate = {2025-10-20},
    year = {1978},
	date = {1978-01-01},
	langid = {english},
}

@article{akaike_new_1974,
	title = {A new look at the statistical model identification},
	volume = {19},
	rights = {https://ieeexplore.ieee.org/Xplorehelp/downloads/license-information/{IEEE}.html},
	issn = {0018-9286},
	url = {http://ieeexplore.ieee.org/document/1100705/},
	doi = {10.1109/TAC.1974.1100705},
	pages = {716--723},
	number = {6},
	journal = {{IEEE} Transactions on Automatic Control},
	shortjournal = {{IEEE} Trans. Automat. Contr.},
	author = {Akaike, H.},
	urldate = {2025-10-20},
    year = {1974},
	date = {1974-12},
	langid = {english},
}

@book{burnham_model_2010,
	location = {New York, {NY}},
	edition = {2. ed},
	title = {Model selection and multimodel inference: a practical information-theoretic approach},
	isbn = {978-0-387-22456-5 978-1-4419-2973-0},
	shorttitle = {Model selection and multimodel inference},
	pagetotal = {488},
	publisher = {Springer},
	author = {Burnham, Kenneth P. and Anderson, David Ray},
    year = {2010},
	date = {2010},
}

@book{kutner_applied_2005,
	location = {Boston, Mass.},
	edition = {5. ed},
	title = {Applied linear statistical models},
	isbn = {978-0-07-238688-2},
	series = {{McGraw}-Hill/Irwin series Operations and decision sciences},
	pagetotal = {1396},
	publisher = {{McGraw}-Hill Irwin},
	editor = {Kutner, Michael H.},
    year = {2005},
	date = {2005},
}

@article{wherry_new_1931,
	title = {A New Formula for Predicting the Shrinkage of the Coefficient of Multiple Correlation},
	volume = {2},
	issn = {0003-4851},
	url = {http://projecteuclid.org/euclid.aoms/1177732951},
	doi = {10.1214/aoms/1177732951},
	pages = {440--457},
	number = {4},
	journal = {The Annals of Mathematical Statistics},
	shortjournal = {Ann. Math. Statist.},
	author = {Wherry, R. J.},
	urldate = {2025-10-20},
    year = {1931},
	date = {1931-11},
	langid = {english},
}

@article{yang_comparison_2012,
	title = {Comparison of six particle size distribution models on the goodness-of-fit to particulate matter sampled from animal buildings},
	volume = {62},
	issn = {1096-2247, 2162-2906},
	url = {https://www.tandfonline.com/doi/full/10.1080/10962247.2012.671148},
	doi = {10.1080/10962247.2012.671148},
	pages = {725--735},
	number = {6},
	journal = {Journal of the Air \& Waste Management Association},
	shortjournal = {Journal of the Air \& Waste Management Association},
	author = {Yang, Xufei and Lee, Jongmin and Barker, Douglas E. and Wang, Xinlei and Zhang, Yuanhui},
	urldate = {2025-10-20},
    year = {2012},
	date = {2012-06},
	langid = {english},
}

@article{gao_particle_2023,
	title = {Particle Size Distributions and Extinction Coefficients of Aerosol Particles in Land Battlefield Environments},
	volume = {15},
	issn = {2072-4292},
	url = {https://www.mdpi.com/2072-4292/15/20/5038},
	doi = {10.3390/rs15205038},
	pages = {5038},
	number = {20},
	journal = {Remote Sensing},
	shortjournal = {Remote Sensing},
	author = {Gao, Lijuan and Chen, Huimin and Chen, Guang and Deng, Jiahao},
	urldate = {2025-10-20},
    year = {2023},
	date = {2023-10-20},
	langid = {english},
}

@article{kass_bayes_1995,
	title = {Bayes Factors},
	volume = {90},
	issn = {0162-1459, 1537-274X},
	url = {http://www.tandfonline.com/doi/abs/10.1080/01621459.1995.10476572},
	doi = {10.1080/01621459.1995.10476572},
	pages = {773--795},
	number = {430},
	journal = {Journal of the American Statistical Association},
	shortjournal = {Journal of the American Statistical Association},
	author = {Kass, Robert E. and Raftery, Adrian E.},
	urldate = {2025-10-20},
    year = {1995},
	date = {1995-06},
	langid = {english},
}

@article{schwarz_estimating_1978,
	title = {Estimating the Dimension of a Model},
	volume = {6},
	issn = {0090-5364},
	url = {https://projecteuclid.org/journals/annals-of-statistics/volume-6/issue-2/Estimating-the-Dimension-of-a-Model/10.1214/aos/1176344136.full},
	doi = {10.1214/aos/1176344136},
	number = {2},
	journal = {The Annals of Statistics},
	shortjournal = {Ann. Statist.},
	author = {Schwarz, Gideon},
	urldate = {2025-10-20},
    year = {1978},
	date = {1978-03-01},
}

@article{hussein_meteorological_2006,
	title = {Meteorological dependence of size-fractionated number concentrations of urban aerosol particles},
	volume = {40},
	rights = {https://www.elsevier.com/tdm/userlicense/1.0/},
	issn = {13522310},
	url = {https://linkinghub.elsevier.com/retrieve/pii/S135223100501037X},
	doi = {10.1016/j.atmosenv.2005.10.061},
	pages = {1427--1440},
	number = {8},
	journal = {Atmospheric Environment},
	shortjournal = {Atmospheric Environment},
	author = {Hussein, Tareq and Karppinen, Ari and Kukkonen, Jaakko and Härkönen, Jari and Aalto, Pasi P. and Hämeri, Kaarle and Kerminen, Veli-Matti and Kulmala, Markku},
	urldate = {2025-12-15},
    year = {2006},
	date = {2006-03},
	langid = {english},
}

@article{hussein_urban_2004,
	title = {Urban aerosol number size distributions},
	volume = {4},
	rights = {https://creativecommons.org/licenses/by-nc-sa/2.5/},
	issn = {1680-7324},
	url = {https://acp.copernicus.org/articles/4/391/2004/},
	doi = {10.5194/acp-4-391-2004},
	pages = {391--411},
	number = {2},
	journal = {Atmospheric Chemistry and Physics},
	shortjournal = {Atmos. Chem. Phys.},
	author = {Hussein, T. and Puustinen, A. and Aalto, P. P. and Mäkelä, J. M. and Hämeri, K. and Kulmala, M.},
	urldate = {2025-12-15},
    year = {2004},
	date = {2004-03-01},
	langid = {english},
}

@article{aalto_physical_2001,
	title = {Physical characterization of aerosol particles during nucleation events},
	volume = {53},
	issn = {1600-0889, 0280-6509},
	url = {https://b.tellusjournals.se/article/10.3402/tellusb.v53i4.17127/},
	doi = {10.3402/tellusb.v53i4.17127},
	pages = {344},
	number = {4},
	journal = {Tellus B: Chemical and Physical Meteorology},
	author = {Aalto, Pasi and Hämeri, Kaarle and Becker, Edo and Weber, Rodney and Salm, Jaan and Mäkelä, Jyrki M. and Hoell, Claudia and O’dowd, Colin D. and Karlsson, Hans and Hansson, Hans-Christen and Väkevä, Minna and Koponen, Ismo K. and Buzorius, Gintautas and Kulmala, Markku},
	urldate = {2025-12-15},
    year = {2001},
	date = {2001-01-01},
}

@article{kyro_trends_2014,
	title = {Trends in new particle formation in eastern Lapland, Finland: effect of decreasing sulfur emissions from Kola Peninsula},
	volume = {14},
	rights = {https://creativecommons.org/licenses/by/3.0/},
	issn = {1680-7324},
	url = {https://acp.copernicus.org/articles/14/4383/2014/},
	doi = {10.5194/acp-14-4383-2014},
	shorttitle = {Trends in new particle formation in eastern Lapland, Finland},
	pages = {4383--4396},
	number = {9},
	journal = {Atmospheric Chemistry and Physics},
	shortjournal = {Atmos. Chem. Phys.},
	author = {Kyrö, E.-M. and Väänänen, R. and Kerminen, V.-M. and Virkkula, A. and Petäjä, T. and Asmi, A. and Dal Maso, M. and Nieminen, T. and Juhola, S. and Shcherbinin, A. and Riipinen, I. and Lehtipalo, K. and Keronen, P. and Aalto, P. P. and Hari, P. and Kulmala, M.},
	urldate = {2026-03-25},
    year = {2014},
	date = {2014-05-05},
	langid = {english},
}

@article{ahti_vegetation_1968,
	title = {Vegetation zones and their sections in northwestern Europe},
	volume = {5},
	issn = {00033847, 17972442},
	url = {http://www.jstor.org/stable/23724233},
	pages = {169--211},
	number = {3},
	journal = {Annales Botanici Fennici},
	publisher = {Finnish Zoological and Botanical Publishing Board},
	author = {Ahti, Teuvo and Hämet-Ahti, Leena and Jalas, Jaakko},
	urldate = {2026-03-25},
    year = {1968},
	date = {1968},
}

@incollection{hari_station_2013,
	location = {Dordrecht},
	title = {Station for Measuring Ecosystem-Atmosphere Relations: {SMEAR}},
	isbn = {978-94-007-5602-1 978-94-007-5603-8},
	url = {http://link.springer.com/10.1007/978-94-007-5603-8_9},
	doi = {10.1007/978-94-007-5603-8_9},
	shorttitle = {Station for Measuring Ecosystem-Atmosphere Relations},
	pages = {471--487},
	booktitle = {Physical and Physiological Forest Ecology},
	publisher = {Springer Netherlands},
	author = {Hari, Pertti and Nikinmaa, Eero and Pohja, Toivo and Siivola, Erkki and Bäck, Jaana and Vesala, Timo and Kulmala, Markku},
	editor = {Hari, Pertti and Heliövaara, Kari and Kulmala, Liisa},
	urldate = {2026-03-25},
    year = {2013},
	date = {2013},
	langid = {english},
}

@article{kulmala_formation_2004,
	title = {Formation and growth rates of ultrafine atmospheric particles: a review of observations},
	volume = {35},
	rights = {https://www.elsevier.com/tdm/userlicense/1.0/},
	issn = {00218502},
	url = {https://linkinghub.elsevier.com/retrieve/pii/S0021850203004373},
	doi = {10.1016/j.jaerosci.2003.10.003},
	shorttitle = {Formation and growth rates of ultrafine atmospheric particles},
	pages = {143--176},
	number = {2},
	journal = {Journal of Aerosol Science},
	shortjournal = {Journal of Aerosol Science},
	author = {Kulmala, M. and Vehkamäki, H. and Petäjä, T. and Dal Maso, M. and Lauri, A. and Kerminen, V.-M. and Birmili, W. and {McMurry}, P.H.},
	urldate = {2026-03-25},
    year = {2004},
	date = {2004-03},
	langid = {english},
}

@article{hari2005,
  title={Station for Measuring Ecosystem-Atmosphere Relations (SMEAR II)},
  author={Hari, P and  Kulmala, M},
  journal={Boreal Environment Research},
  volume={10},
  number={5},
  pages={345-322},
  year={2005}
}

@article{Uusitalo2021,
  title={Occurrence of new particle formation events in Siberian and Finnish boreal forest},
  author={Uusitalo, Helmi and Kontkanen, Jenni and Ylivinkka, Ilona and Ezhova, Ekaterina and Demakova, Anastasiia and Arshinov, Mikhail and Belan, Boris Denisovich and Davydov, Denis and Ma, Nan and Pet{\"a}j{\"a}, Tuukka and others},
  journal={Atmospheric Chemistry and Physics Discussions},
  volume={2021},
  pages={1--21},
  year={2021},
  publisher={G{\"o}ttingen, Germany}
}

@article{Zaidan2018,
  title = {Predicting atmospheric particle formation days by Bayesian classification of the time series features},
  volume = {70},
  ISSN = {0280-6509},
  url = {http://dx.doi.org/10.1080/16000889.2018.1530031},
  DOI = {10.1080/16000889.2018.1530031},
  number = {1},
  journal = {Tellus B: Chemical and Physical Meteorology},
  publisher = {Stockholm University Press},
  author = {Zaidan,  M. A. and Haapasilta,  V. and Relan,  R. and Junninen,  H. and Aalto,  P. P. and Kulmala,  M. and Laurson,  L. and Foster,  A. S.},
  year = {2018},
  month = Jan,
  pages = {1530031}
}

@article{Karpatne2019,
  title = {Machine Learning for the Geosciences: Challenges and Opportunities},
  volume = {31},
  ISSN = {2326-3865},
  url = {http://dx.doi.org/10.1109/TKDE.2018.2861006},
  DOI = {10.1109/tkde.2018.2861006},
  number = {8},
  journal = {IEEE Transactions on Knowledge and Data Engineering},
  publisher = {Institute of Electrical and Electronics Engineers (IEEE)},
  author = {Karpatne,  Anuj and Ebert-Uphoff,  Imme and Ravela,  Sai and Babaie,  Hassan Ali and Kumar,  Vipin},
  year = {2019},
  month = Aug,
  pages = {1544–1554}
}

@article{Du2021,
  title = {Deep Air Quality Forecasting Using Hybrid Deep Learning Framework},
  volume = {33},
  ISSN = {2326-3865},
  url = {http://dx.doi.org/10.1109/TKDE.2019.2954510},
  DOI = {10.1109/tkde.2019.2954510},
  number = {6},
  journal = {IEEE Transactions on Knowledge and Data Engineering},
  publisher = {Institute of Electrical and Electronics Engineers (IEEE)},
  author = {Du,  Shengdong and Li,  Tianrui and Yang,  Yan and Horng,  Shi-Jinn},
  year = {2021},
  month = {June},
  pages = {2412–2424}
}

@article{harni_source_2024,
	title = {Source apportionment of particle number size distribution at the street canyon and urban background sites},
	volume = {24},
	rights = {https://creativecommons.org/licenses/by/4.0/},
	issn = {1680-7324},
	url = {https://acp.copernicus.org/articles/24/12143/2024/},
	doi = {10.5194/acp-24-12143-2024},
	pages = {12143--12160},
	number = {21},
	journal = {Atmospheric Chemistry and Physics},
	shortjournal = {Atmos. Chem. Phys.},
	author = {Harni, Sami D. and Aurela, Minna and Saarikoski, Sanna and Niemi, Jarkko V. and Portin, Harri and Manninen, Hanna and Leinonen, Ville and Aalto, Pasi and Hopke, Phil K. and Petäjä, Tuukka and Rönkkö, Topi and Timonen, Hilkka},
	urldate = {2026-05-19},
	date = {2024-10-30},
    year = {2024},
	langid = {english},
}

@article{de_jesus_ultrafine_2019,
	title = {Ultrafine particles and {PM}2.5 in the air of cities around the world: Are they representative of each other?},
	volume = {129},
	issn = {01604120},
	url = {https://linkinghub.elsevier.com/retrieve/pii/S0160412019311110},
	doi = {10.1016/j.envint.2019.05.021},
	shorttitle = {Ultrafine particles and {PM}2.5 in the air of cities around the world},
	pages = {118--135},
	journal = {Environment International},
	shortjournal = {Environment International},
	author = {De Jesus, Alma Lorelei and Rahman, Md Mahmudur and Mazaheri, Mandana and Thompson, Helen and Knibbs, Luke D. and Jeong, Cheol and Evans, Greg and Nei, Wei and Ding, Aijun and Qiao, Liping and Li, Li and Portin, Harri and Niemi, Jarkko V. and Timonen, Hilkka and Luoma, Krista and Petäjä, Tuukka and Kulmala, Markku and Kowalski, Michal and Peters, Annette and Cyrys, Josef and Ferrero, Luca and Manigrasso, Maurizio and Avino, Pasquale and Buonano, Giorgio and Reche, Cristina and Querol, Xavier and Beddows, David and Harrison, Roy M. and Sowlat, Mohammad H. and Sioutas, Constantinos and Morawska, Lidia},
	urldate = {2026-05-19},
	date = {2019-08},
    year = {2019},
	langid = {english},
}

@book{burnham_model_2004,
	location = {New York, {NY}},
	title = {Model Selection and Multimodel Inference},
	rights = {http://www.springer.com/tdm},
	isbn = {978-0-387-95364-9},
	url = {http://link.springer.com/10.1007/b97636},
	doi = {10.1007/b97636},
	publisher = {Springer New York},
	editor = {Burnham, Kenneth P. and Anderson, David R.},
	urldate = {2026-05-21},
	date = {2004},
    year = {2004},
	langid = {english},
}

@article{sulo_long-term_2021,
	title = {Long-term measurement of sub-3 nm particles and their precursor gases in the boreal forest},
	volume = {21},
	rights = {https://creativecommons.org/licenses/by/4.0/},
	issn = {1680-7324},
	url = {https://acp.copernicus.org/articles/21/695/2021/},
	doi = {10.5194/acp-21-695-2021},
	pages = {695--715},
	number = {2},
	journal = {Atmospheric Chemistry and Physics},
	shortjournal = {Atmos. Chem. Phys.},
	author = {Sulo, Juha and Sarnela, Nina and Kontkanen, Jenni and Ahonen, Lauri and Paasonen, Pauli and Laurila, Tiia and Jokinen, Tuija and Kangasluoma, Juha and Junninen, Heikki and Sipilä, Mikko and Petäjä, Tuukka and Kulmala, Markku and Lehtipalo, Katrianne},
	urldate = {2026-05-28},
	date = {2021-01-19},
    year = {2021},
	langid = {english},
}

\end{document}